\documentclass[a4paper]{jpconf}
\usepackage{graphicx,epstopdf}
\usepackage[small,it]{caption}

\begin{document}
\title{From black holes to their progenitors: A full population study in measuring black hole binary
parameters from ringdown signals}

\author{Ioannis Kamaretsos}
\address{School of Physics and Astronomy, Cardiff University, Queens Buildings, CF24 3AA, Cardiff, United Kingdom}
\ead{ioannis.kamaretsos@astro.cf.ac.uk}

\begin{abstract}
A perturbed black hole emits gravitational radiation, usually termed the ringdown signal, whose frequency and time-constant depends on the mass and spin of the black hole. I investigate the case of a binary black hole merger resulting from two initially non-spinning 
black holes of various mass ratios, in quasi-circular orbits. The observed ringdown signal will be determined, among other things, by the black hole's spin-axis orientation with respect to Earth, its sky position and polarization angle - parameters which can take any values in a particular observation. I have carried out a statistical analysis of the effect of these variables, focusing on detection and measurement of the multimode ringdown signals using the reformulated European LISA mission, Next Gravitational-Wave Observatory, NGO, the third generation ground-based observatory, Einstein Telescope and the advanced era detector, aLIGO. To the extent possible I have discussed the effect of these results on plausible event rates, as well as astrophysical implications concerning the formation and growth of supermassive and intermediate mass black holes.   
\end{abstract}

\section{Introduction}
Astrophysical observations to date have provided sturdy evidence that black holes (BHs) may exist and play an important role in many physical processes \cite{Nandra:2009br, Somerville:2008ch, AmaroSeoane:2009ui}. With direct evidence still lacking, it is expected that  observation of gravitational waves (GWs) from merged BHs will not only provide indisputable evidence for the existence of BHs, but also the ability to extract accurate information about the progenitor system and the BH. 

The ringdown radiation consists of a superposition of, in principle, an infinite number of essentially damped sinusoids, termed quasi-normal modes. Their frequencies and time-constants depend only on the mass and spin of the BH -- a consequence of the {\em no-hair} theorem \cite{BCW05}.
In a recent work \cite{PhysRevD.85.024018} we have argued that the amplitude terms of the various quasi-normal modes encode important information about the origin of the perturbation that caused them, such as the component masses of the progenitor binary. This allows performing parameter estimation on the system from the strong-field regime, as opposed to using the inspiral phase. However, in that study, as well as in previous studies of parameter estimation from ringdown signals \cite{BCW05,Berti:2007a}, only a small region of the parameter space was explored. In a realistic scenario, a BH ringdown signal can have any sky location and polarization. All the while, it could have originated from a BH with any spin-axis orientation with respect to Earth. These variables have a significant effect on the observed ringdown signal and a direct impact on the science we can achieve by observing GWs from merged BHs.

In the present study, I have investigated in detail how these angular parameters affect the detection and measurement of the ringdown signals. To this end, I have varied the angular parameters over their full range, thereby considering a large population of BH mergers. I have considered supermassive BHs (SMBHs of mass $\geq 10^6M_\odot$) visible in NGO and intermediate mass BHs (IMBHs of mass $\sim10^3M_\odot$) observable in ET and aLIGO. I have computed the probability distribution functions of signal to noise ratios (SNR), as well as measurement errors of a chosen set of parameters, for a wide range of the BH mass and for mass ratios between 2 and 20. Finally, I translate these probabilities to proportions of observed events in NGO and ET that will yield parameter errors below certain thresholds and discuss how observations of ringdown signals could help in dealing with open questions on the existence and history of SMBHs and IMBHs. 
\vspace{-0.19in}

\begin{figure}[ht]
\begin{minipage}[b]{0.5\linewidth}
\includegraphics[scale=0.3]{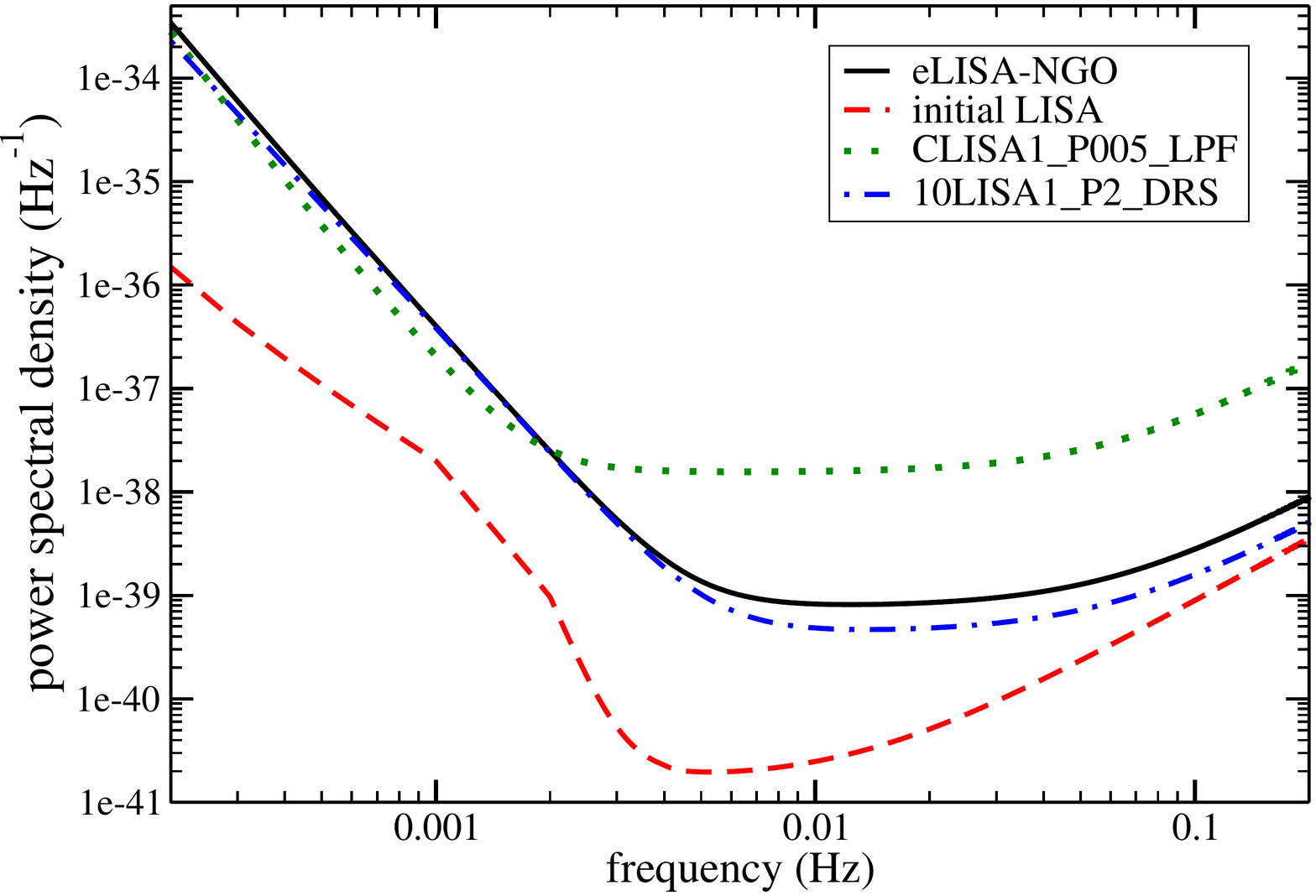}
\caption{Noise power spectral densities for various proposed configurations of LISA-like space detectors. The dashed red line corresponds to the original LISA mission, while the rest of them refer to European designs of LISA. In all cases, the galactic binary white dwarf confusion noise \cite{Nelemans:2003ha, VanDenBroeck:2010fp} is included, which has a negligible visible effect on the newer LISA curves though, due to their, almost two orders of magnitude worse sensitivity. Additionally, a low frequency cut-off - not shown - was induced at $5\times10^{-5}Hz$. In this study, I am using the latest arrangement for the European mission of LISA, labeled eLISA-NGO in this graph.}
\label{fig:PSDs}
\end{minipage}
\hspace{5pt}
\begin{minipage}[b]{0.5\linewidth}
\includegraphics[scale=0.3]{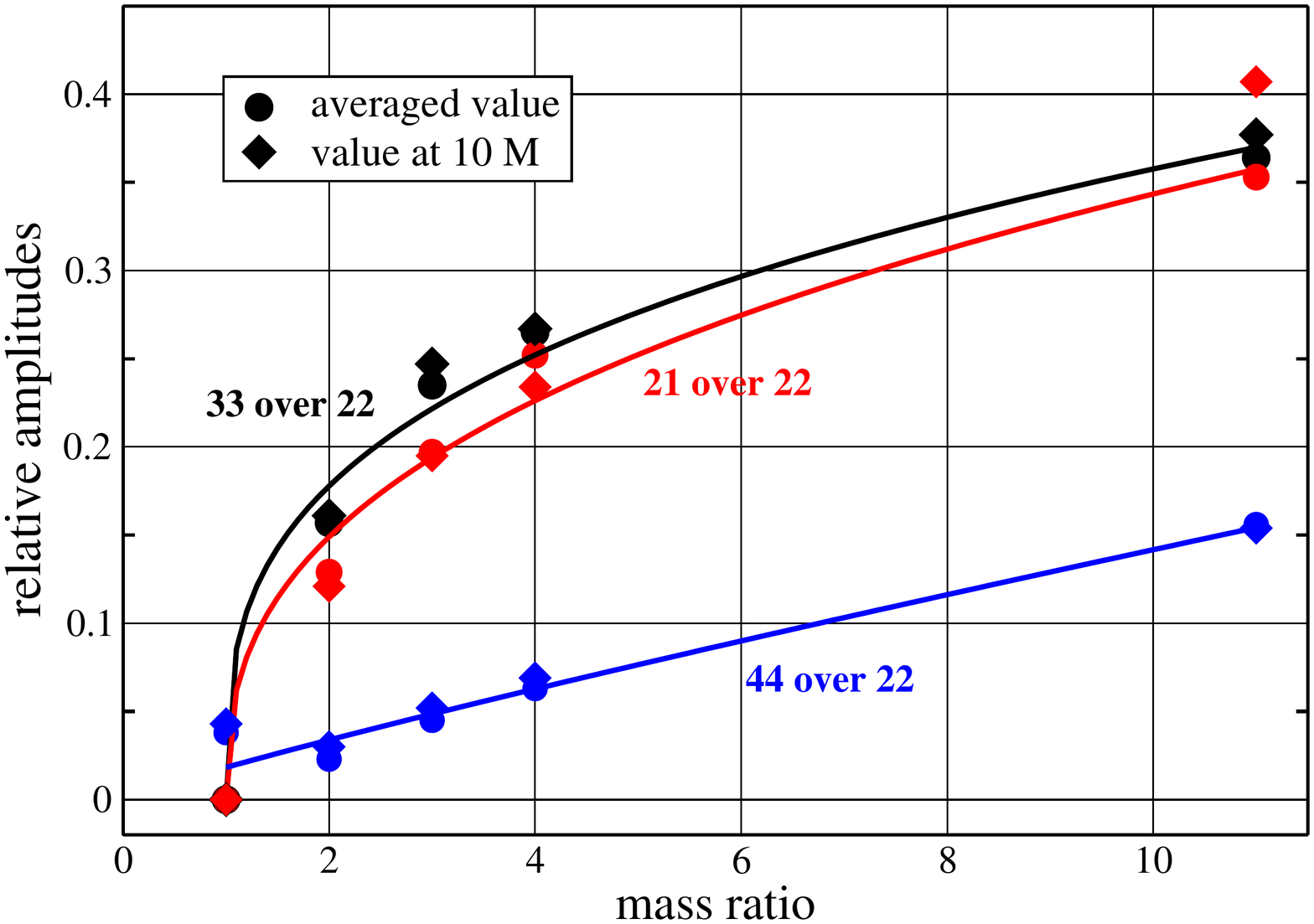}
\caption{This plot shows the relative amplitudes of modes $(3,3)$, $(2,1)$ and $(4,4)$ over $(2,2)$ as a function of the mass ratio. The diamond points show the values that correspond to a time 10$M$ after the peak luminosity of the $22$ mode in the equal mass case. (see also Fig.3 in \cite{PhysRevD.85.024018}) Fits to these points were used in this previous work. On the other hand, the circle points, which are used in the present study, were computed by taking into account all the points in the waveform, in a time region starting at the peak luminosity of $22$ and ending 30$M$ later. The solid lines shown, constitute fits to these circle points, given by expressions (\ref{eq:a33})-(\ref{eq:a44}).} 
\label{fig:relampl}
\end{minipage}
\end{figure}
\vspace{-0.05in} 
\section{Full population Analysis}
I have closely followed the procedure of Ref. \cite{PhysRevD.85.024018} for estimating the signal-to-noise ratios and measurement errors. In this Section I will discuss the signal model and the parameter space covered in this study.

In the generic case where we have a network of detectors, we write the response to a ringdown signal as:
\begin{equation}
h^a(t) = \sum_{\ell,m,n\geq0} B_{\ell m n}^a e^{-t/\tau_{\ell m n}}
\cos\left (\omega_{\ell m n} t + \gamma_{\ell m n}^a \right ),
\label{eq:reduced response}
\end{equation}
\vspace{-0.33in} 

where the superscript $a$ is an index denoting the detector in question and $\omega_{\ell m n}$, $\tau_{\ell m n}$ are the frequencies and time-constants of each mode, which are functions of the mass and spin magnitude of the BH. For further reference, see \cite{Berti:2009kk, Kokkotas:1999bd}. In this study I neglect modes with overtone index $n\geq1$, thereby considering only the least damped modes. From now on, the $n$ index ($n$=$0$) is omitted.
The terms $B_{\ell m}$ and $\gamma_{\ell m}$ are the following combinations of the antenna pattern functions $F^a_+$, $F^a_\times$, amplitude factors $\alpha_{\ell m}$ and the angular functions $Y^{\ell m}_+ (\iota)$, $Y^{\ell m}_\times(\iota):$ 
\begin{eqnarray}
B_{\ell m}^a & = & \frac{M \alpha_{\ell m}}{D_{\rm L}} \sqrt{ \left ( F^a_+\, Y^{\ell m}_+ \right )^2  
+ \left (F^a_\times\, Y^{\ell m}_\times \right )^2},
\label{eq:blm}\\
\gamma_{\ell m}^a & = & \phi_{\ell m} + m\, \phi + \tan^{-1} \left[ \frac{F^a_\times\, Y^{\ell m}_\times}
{F^a_+\, Y^{\ell m}_+} \right ].
\label{eq:gammalm}
\end{eqnarray}

Here, $\phi_{\ell m}$ are arbitrary constant phases of each mode. The effective amplitudes $B_{\ell m}$ vary inversely with the luminosity distance and proportionally to the intrinsic amplitudes $\alpha_{\ell m}$ of the modes, which are determined by the numerical simulations. The angular functions $Y^{\ell m}_{+,\times}(\iota)$ are the following combinations of the spin-weighted spherical harmonics \cite{Berti:2007a}:
\begin{eqnarray}
Y^{\ell m}_+(\iota) & \equiv & {_{-2}Y^{\ell m}}(\iota,0) + (-1)^\ell\,  {_{-2}Y^{\ell -m}}(\iota,0),\nonumber\\
Y^{\ell m}_\times(\iota) & \equiv & {_{-2}Y^{\ell m}}(\iota,0) - (-1)^\ell\,  {_{-2}Y^{\ell -m}}(\iota,0).
\end{eqnarray}

The antenna pattern functions are functions of the sky location coordinates, $\theta$ and $\phi$ and the polarization angle $\psi$, that is, $F^a_+ (\theta,\varphi,\psi)$, $F^a_\times (\theta,\varphi,\psi)$. The spheroidal harmonics are angular functions of the inclination angle, $\iota$ and the azimuth angle $\phi$. The first refers to the angle formed by the BH's spin angular momentum and the line-of-sight, while the latter is the azimuth angle defined in a non-rotating frame fixed to the BH.

\subsection{Chosen waveform}
The ringdown waveform used is of the form described by Eqs.\,(\ref{eq:reduced response})-(\ref{eq:gammalm}). 
It is a signal comprised of four modes, with mode indices $(\ell,m,n)=(2,2,0), (3,3,0), (2,1,0), (4,4,0)$. Our choice is based on the ordering of the various modes according to power output, as determined from numerical simulations of non-spinning unequal mass binaries \cite{PhysRevD.85.024018,Baker:2008d78,BCW05}.

Numerical simulations of merging black-hole binaries were performed using the BAM code \cite{Brugmann:2008zz,Husa:2007hp}, so as to obtain the amplitudes, $\alpha_{\ell m}$ of the various modes in Eq.(\ref{eq:blm}), as well as their frequencies and time-constants (see also Table I of \cite{PhysRevD.85.024018}). These simulations involve the case of \emph{initially non-spinning} BHs in quasi-circular orbits and for different mass ratios of the binary. For the mass ratios, $q = \{1,2,3,4\}$ the simulation results were first presented in \cite{Berti:2007b, Hannam:2007ik, Hannam:2010ec}, while an additional simulation of a $q=11$ binary was carried out in \cite{PhysRevD.85.024018}.

The amplitude terms $\alpha_{\ell m}$ in Eq. \ref{eq:blm}, are given by the expressions:\vspace{-0.22in}

\begin{eqnarray}
\alpha_{22}(q) &=  & 0.25\, e^{-q/7.5},\label{eq:a22}\\
\alpha_{33}(q) & = & 0.18\, \alpha_{22}(q)\, (q-1)^{0.32},\label{eq:a33}\\
\alpha_{21}(q) & = & 0.15\, \alpha_{22}(q)\, (q-1)^{0.38},\label{eq:a21}\\
\alpha_{44}(q) & = & 0.018\, \alpha_{22}(q)\, q^{0.89}.\label{eq:a44}
\end{eqnarray}
These constitute fits, that were produced by fitting the merger-ringdown part of the numerical simulations data, taking into account all the different mass ratios for which these were performed. All points in a time region beginning at the peak luminosity of $22$ and ending 30$M$ later were considered, for each mass ratio. As opposed to the method that was applied in our previous work \cite{PhysRevD.85.024018}, where the relative amplitude values at 10$M$ were used, this approach is expected to be more robust and to average out any numerical noise that might be present in this part of the waveform. Nevertheless, as can be seen from Fig. \ref{fig:relampl}, these two methods do not give very different results. Note also that the above fitting functions, as well as the mode frequencies and time-constants, may be less accurate in the higher mass ratio values of around 20, where extrapolation has been performed.

\subsection{The simulations}
\label{subsec:simul}
In Ref. \cite{PhysRevD.85.024018} we had ignored the effect of the various angles $\{\theta,\,\varphi,\,\psi,\,\iota\}$ on the quasi-normal mode spectrum and their impact on the detection and measurement of ringdown signals. To assess this effect, the aforementioned analysis was repeated by varying the angular parameters $\{\theta,\,\varphi,\,\psi,\,\iota\}$. Specifically, six uniformly spaced values were chosen for these angles. This results in $6^4=1296$ distinct relative orientations between the detector and the BH and its spin axis. Additionally, a couple of simulations with eight uniformly spaced values were performed, that is $8^4=4096$ configurations, to allow the comparison. Hence, it was decided that six values in each parameter was acceptable in capturing the behaviour of the observable quantities.  

The values in the polarization angle, $\psi$ and the azimuth sky location angle, $\varphi$ were linearly sampled in the ranges $[0, \pi]$ and $[0, 2\pi]$ respectively.
Whereas, the values in the inclination angle $\iota$ and sky position, $\theta$ - which range from 0 to $\pi$ - are deduced from the uniformly spaced values of $\cos(\iota)$ and $\cos(\theta)$ in the range $[-1, +1]$. Note that this excludes configurations of optimally oriented binaries, that is of $\iota=0$ and $\iota=\pi$.

\vspace{-0.04in} 
\begin{figure*}
\begin{tabular}{|c|c|c|}
\hline
\includegraphics[width=0.31\textwidth]{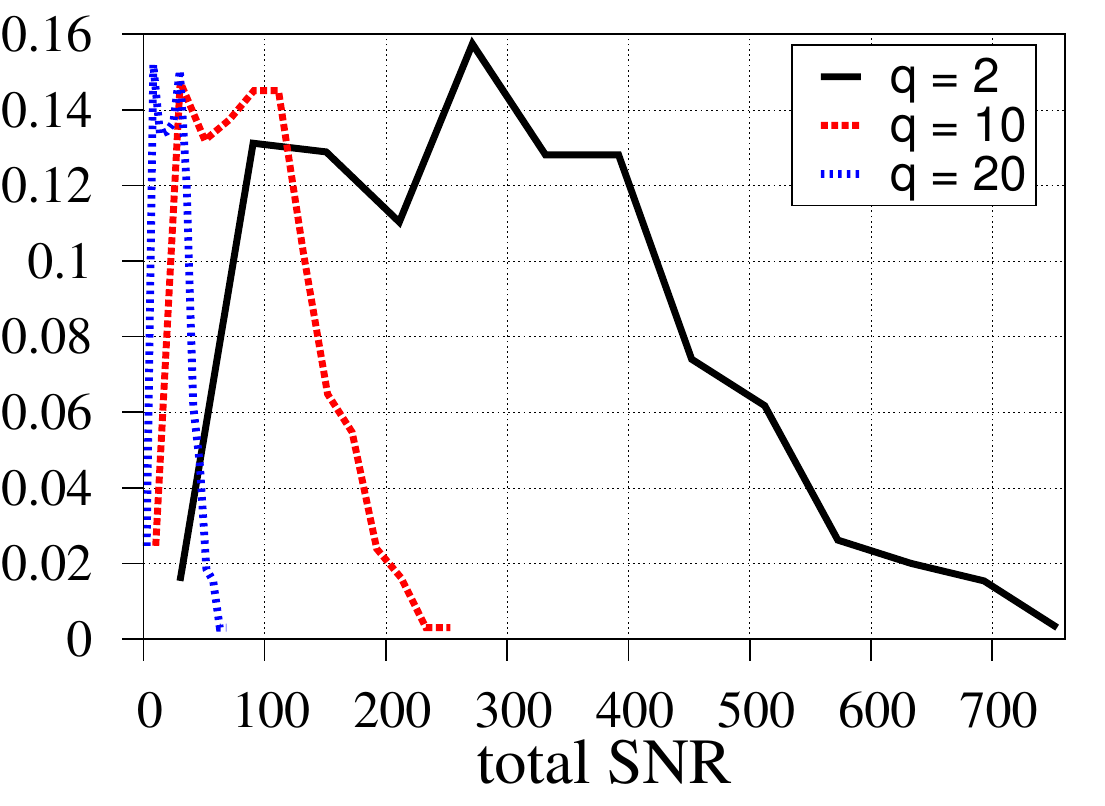}   &
\includegraphics[width=0.31\textwidth]{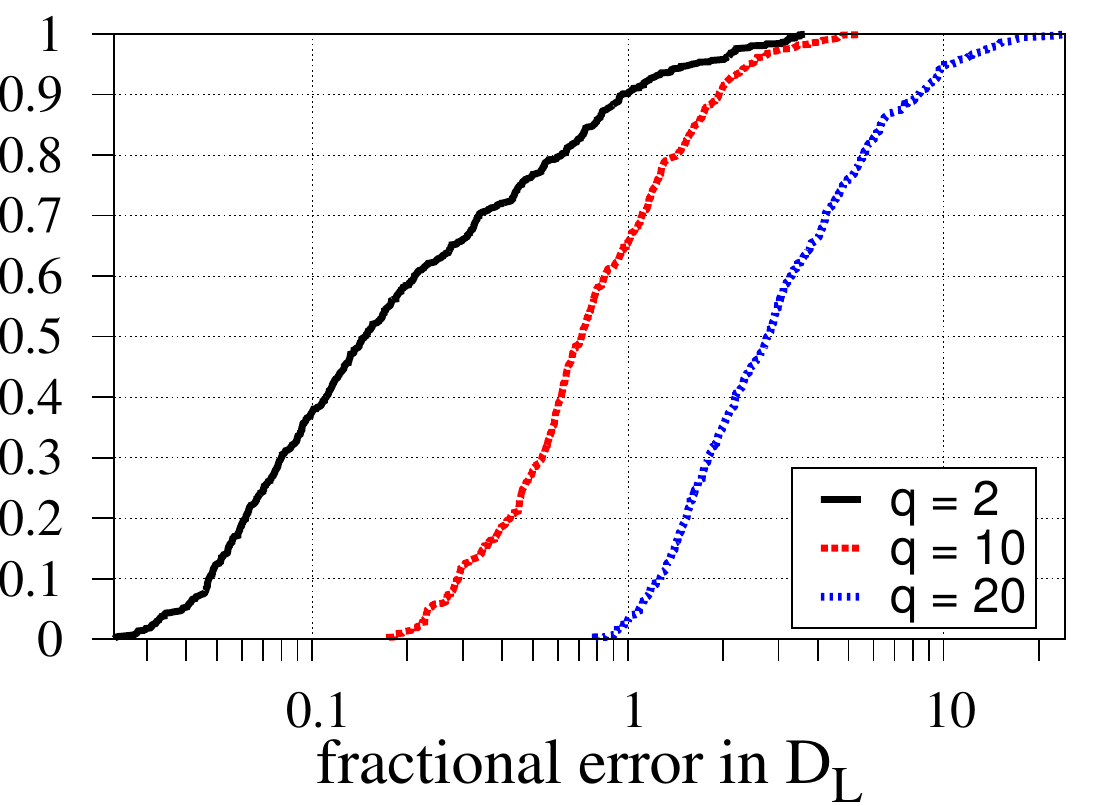}  &
\includegraphics[width=0.31\textwidth]{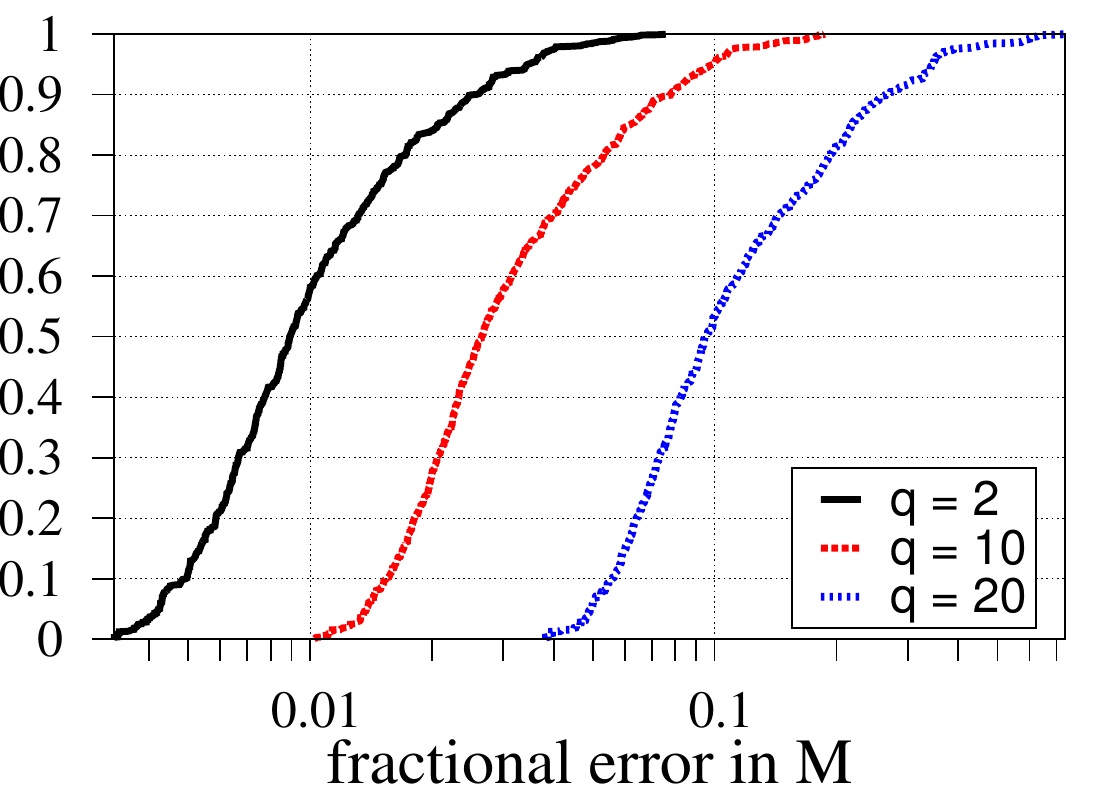}   \\
\hline
\includegraphics[width=0.31\textwidth]{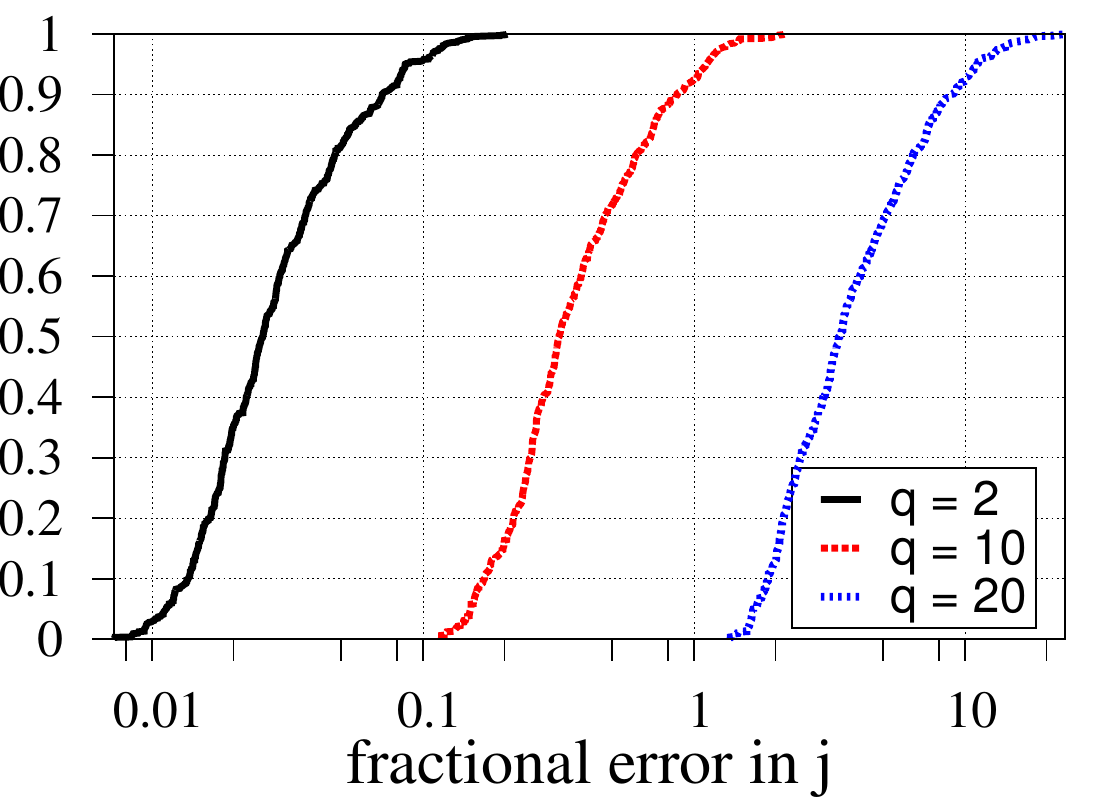}  &
\includegraphics[width=0.31\textwidth]{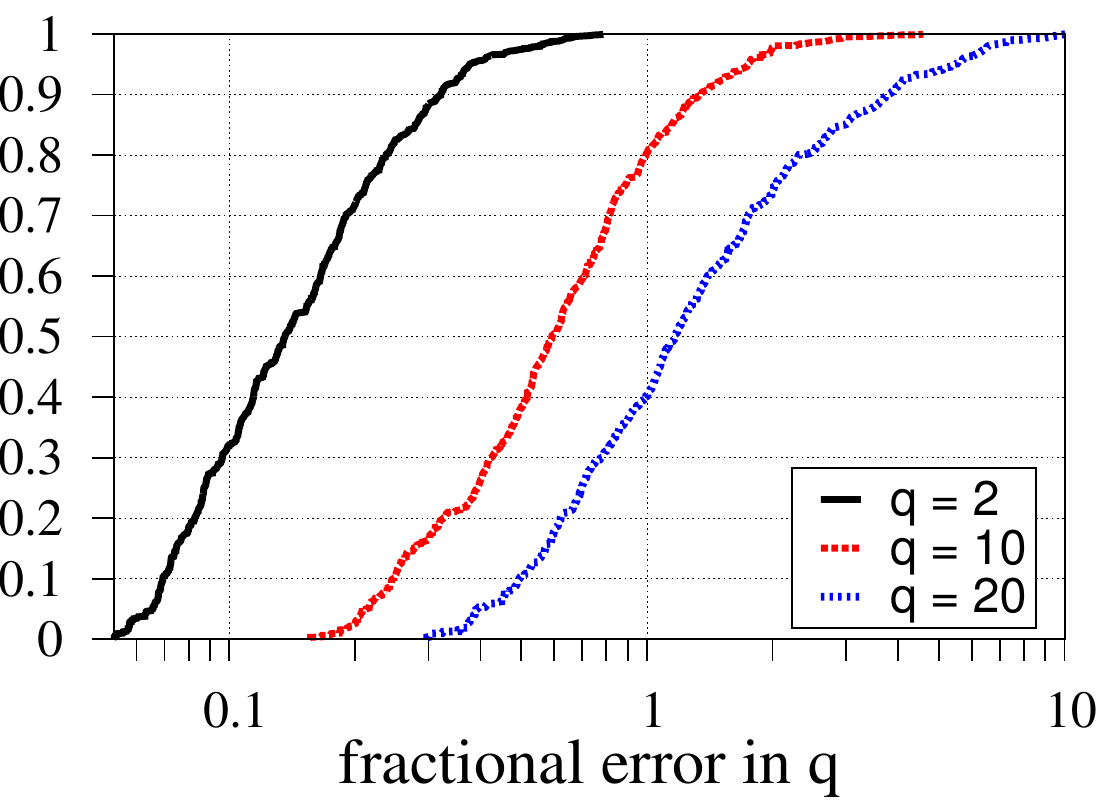} &
\includegraphics[width=0.31\textwidth]{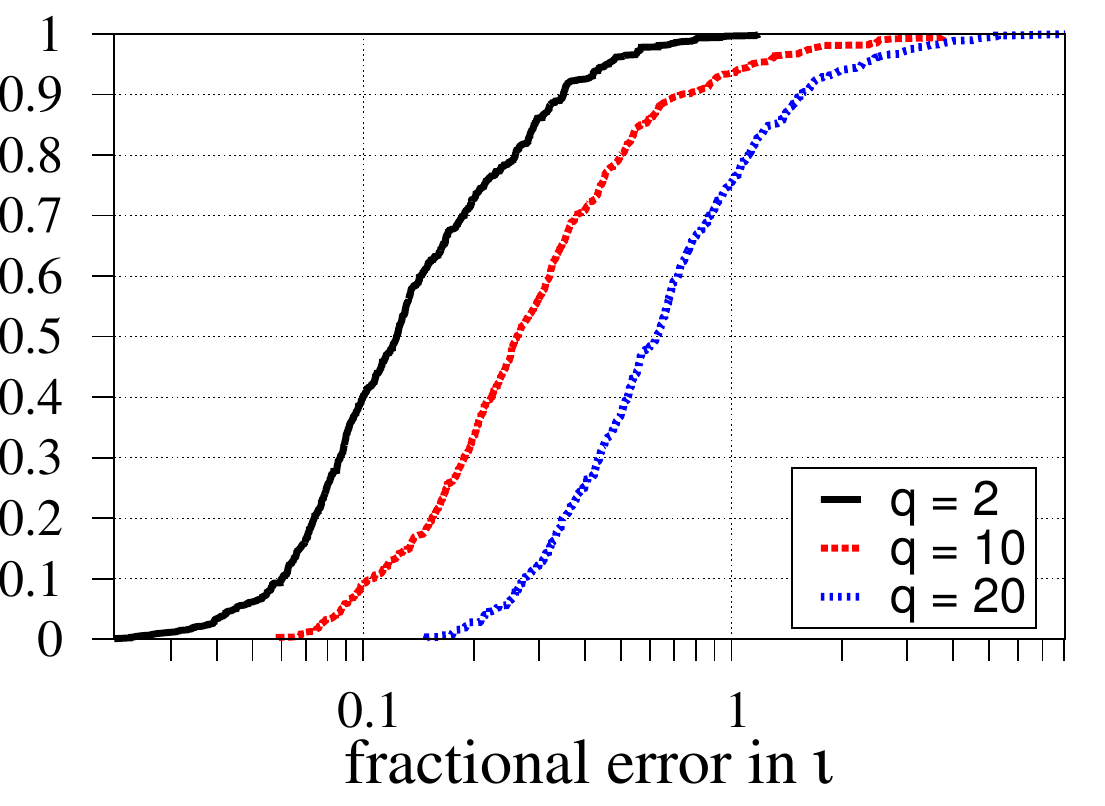}  \\
\hline
\end{tabular}
\caption{Frequency distributions for the total SNR and the measurement errors in NGO, for a $5\times10^6M_{\odot}$ BH situated at a luminosity distance of 6.73 Gpc $(z\simeq1).$ The top left plot shows the probability to have a detector-BH configuration which will yield the SNR shown, while the rest of the plots concern cumulative frequency distributions for the measurement errors. In each graph, the comparison is shown among different mass ratios, $q$, specifically taking the values of 2, 10 and 20. For q=2 - black solid lines - the parameter estimation for the BH mass and spin is outstanding in all configurations. This effectiveness degrades considerably with increasing q. For q of around 10 it is still acceptable, while at q$\simeq$20, all parameters except the mass are very likely to have huge errors, of the order of $200\%$.}
\label{fig:5e6}
\end{figure*}
\vspace{-0.04in} 

\subsection{The parameter space}
The parameter set of the ringdown signal in the case of a non-spinning black hole binary, consists of the following nine parameters: $\{M,\, j,\, q,\, D_{\rm L},\, \theta,\, \varphi,\, \psi,\, \iota,\, \phi\}$. Namely, the mass $M$, the dimensionless Kerr parameter or spin magnitude, $j$ of the BH, the mass ratio $q$ of the progenitor binary, the sky location vector $(D_{\rm L},\,\theta,\,\varphi)$ of the BH with respect to Earth, the polarization angle $\psi$, the inclination angle $\iota$ and the BH azimuth angle $\phi$. Note however, that in this case, the final spin of the BH is directly mapped to the mass ratio of the progenitor binary, therefore $q$ and $j$ are not treated as independent. 

The above parameters are the standard ones, as they pertain to all of the modes. Additional parameters can be introduced, that are characteristic to each mode, such as an initial phase factor $\phi_{\ell  m}$, see Eq.\,(\ref{eq:gammalm}). Therefore, the total number of parameters can increase with the number of modes. We are considering a four mode signal, therefore we have a total of 13 parameters. One thing to note is that the $\ell = m$ modes have a nearly consistent rotational phasing, while the $\ell \neq m$ modes seem to have somewhat distinct associated dynamics, with differentiated amplitude and phasing during the merger process \cite{Baker:2008d78}.  

By virtue of the large number of parameters involved, it was unmanageable to treat the effect of all of them in this analysis. Thus, some of the above mentioned parameters had to be fixed. Specifically, the initial phase angles, $\phi_{\ell  m  0}$ in all the four modes considered, were plainly chosen as zero. In addition, the luminosity distance is chosen to be 6.73 Gpc for NGO, and 1 Gpc for ET and aLIGO. Lastly, the BH azimuth angle, $\phi$ was given the value $\pi/3$. Note that this angle does not have an effect on the SNR, but of course needs to be considered in the waveform.
\vspace{-0.04in} 

\begin{figure*}
\begin{tabular}{|c|c|c|}
\hline
\includegraphics[width=0.31\textwidth]{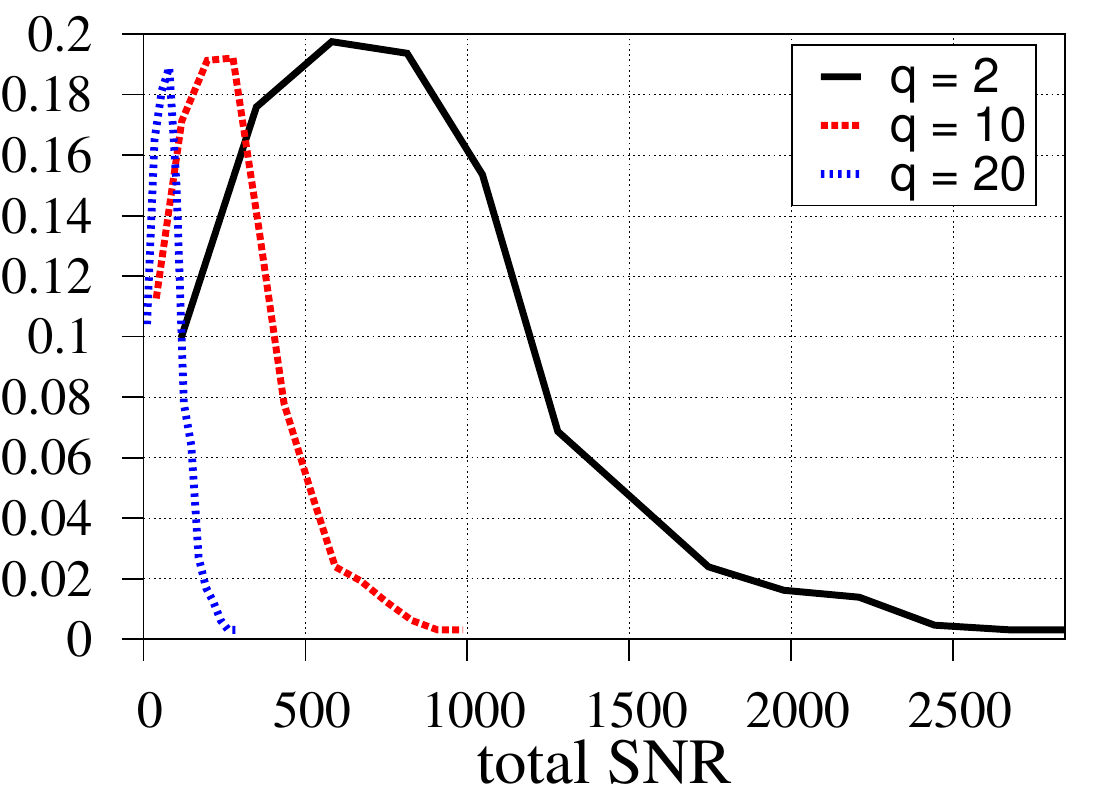}   &
\includegraphics[width=0.31\textwidth]{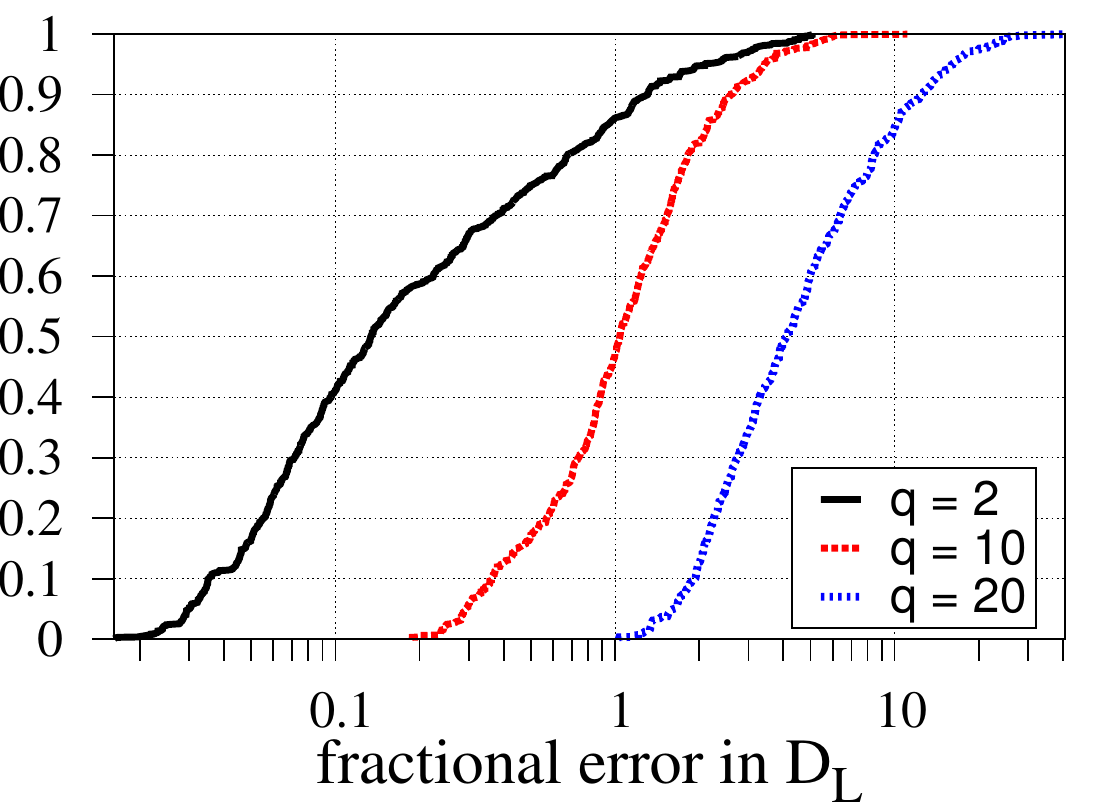}  &
\includegraphics[width=0.31\textwidth]{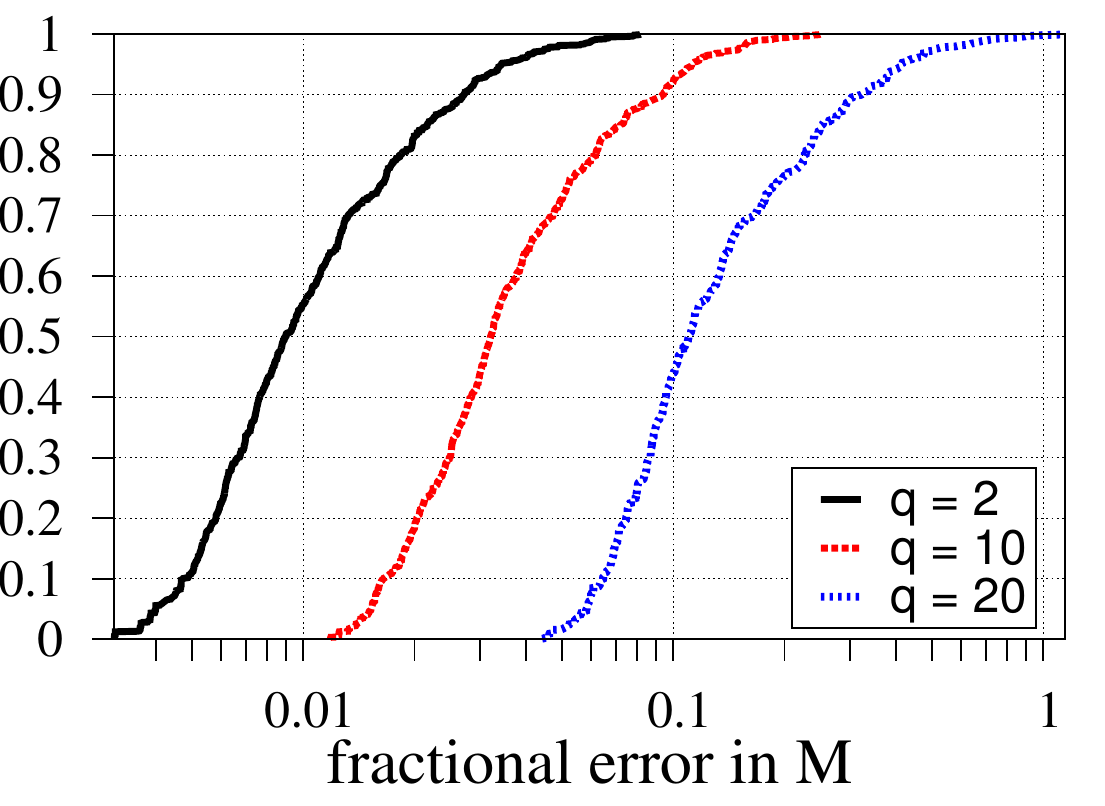}   \\
\hline
\includegraphics[width=0.31\textwidth]{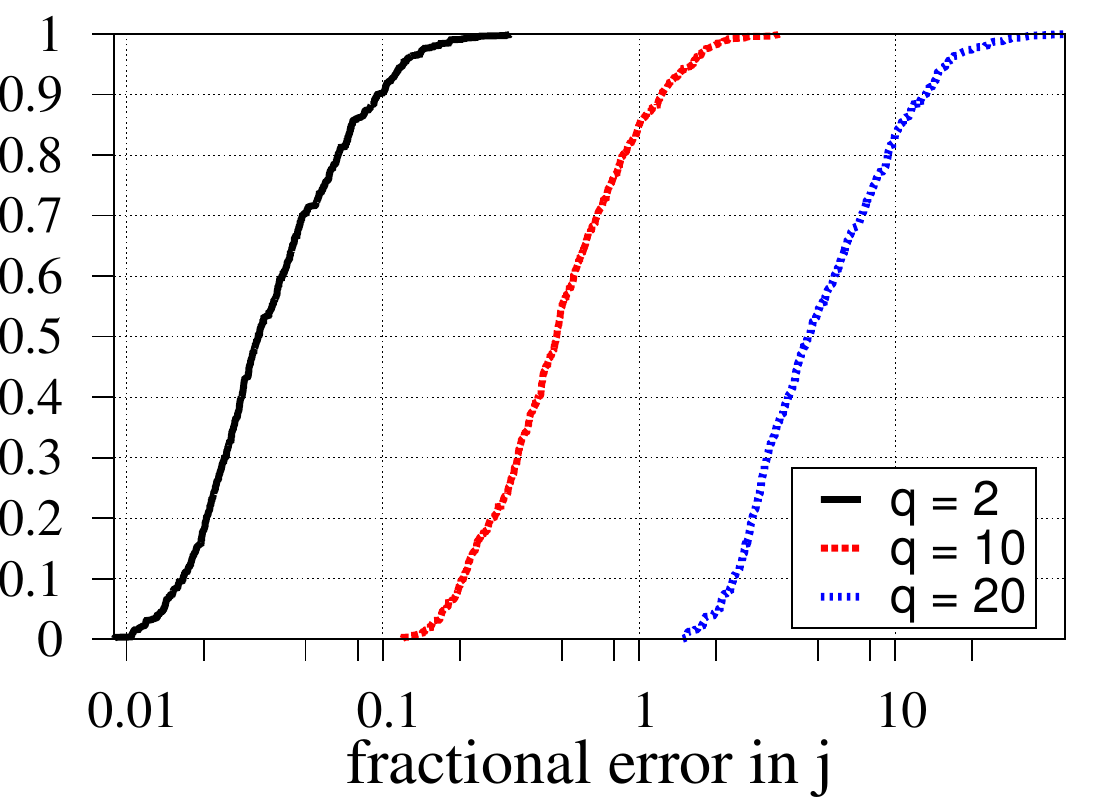}  &
\includegraphics[width=0.31\textwidth]{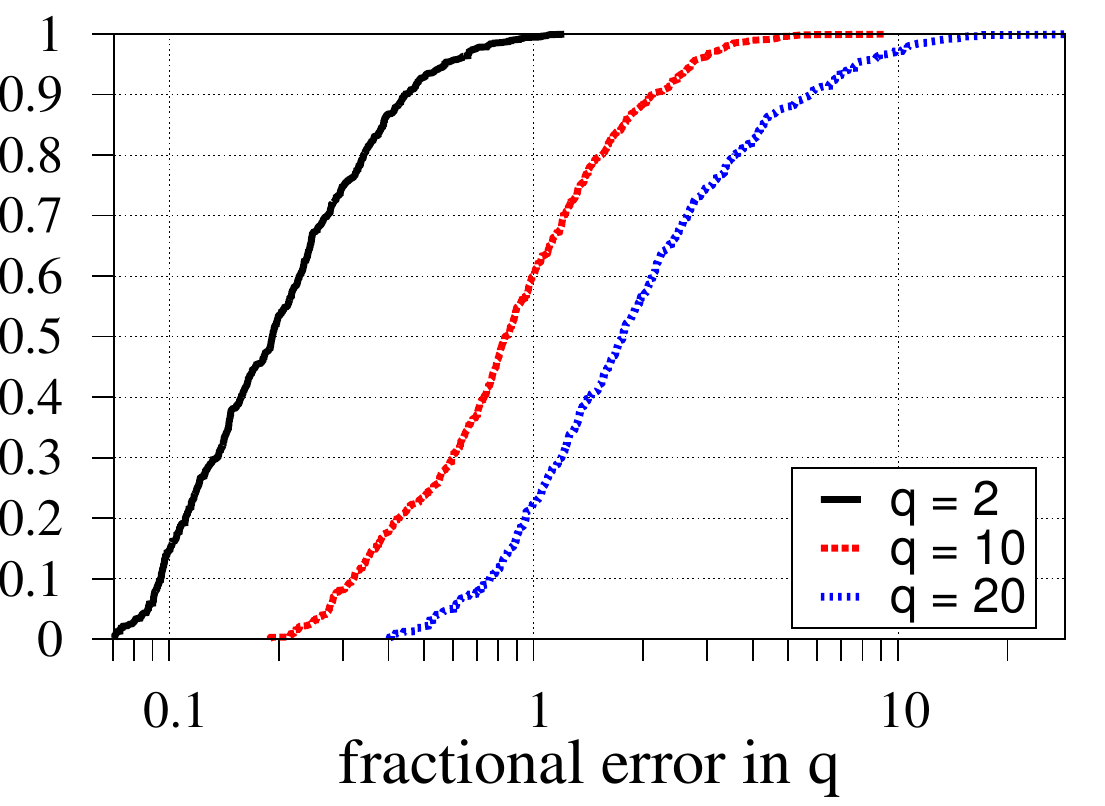} &
\includegraphics[width=0.31\textwidth]{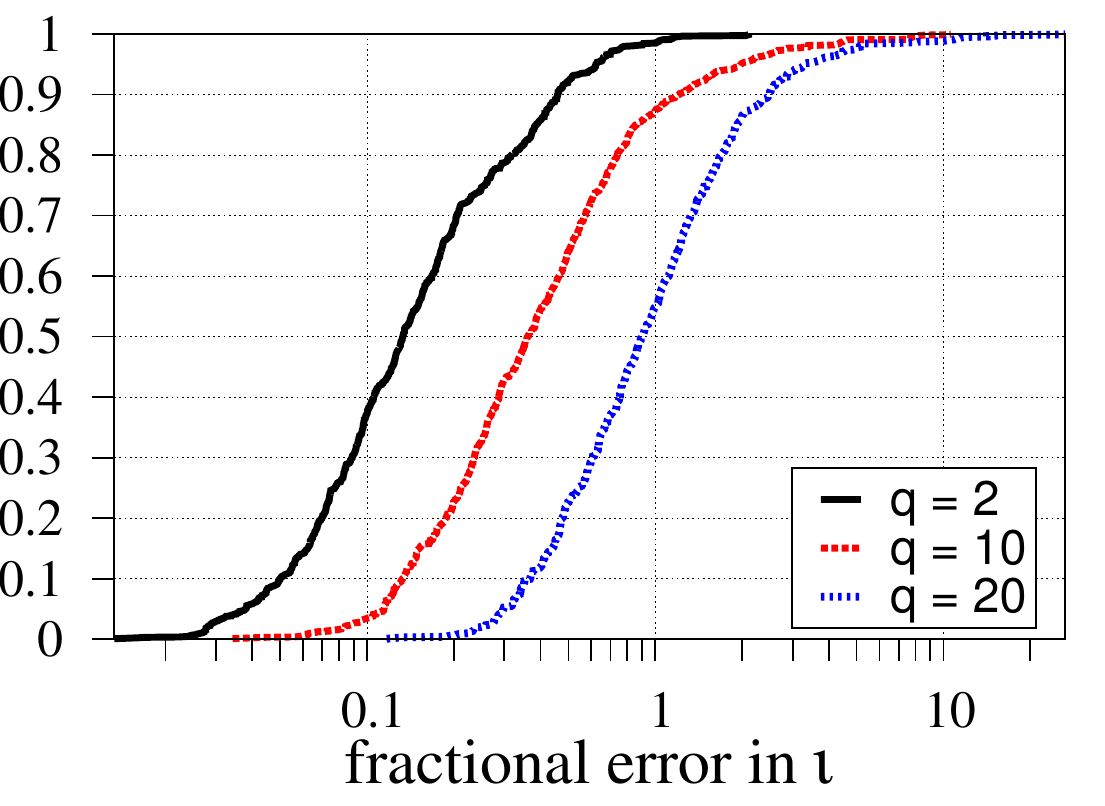}  \\
\hline
\end{tabular}
\caption{Same as in Fig. \ref{fig:5e6}, but concerning a $25\times10^6M_{\odot}$ BH.
This figure concerns a BH 5 times more massive than that of Fig. \ref{fig:5e6}, but some of the results are actually slightly worse, as the multimodal signal's power spectrum is shifted away from the lowest part of NGO's sensitivity curve.}
\label{fig:25e6}
\end{figure*}
\vspace{-0.04in} 

\subsection{Choice of masses and mass ratios}
The BH mass and the binary mass ratio constitute key parameters and the results depend quite strongly on them.
The reasoning behind the choice of mass values is the following. First of all, the low and high mass end is limited by the sensitivity band of the detectors. The existence of BHs in the mass range $10^5-10^6 M_\odot$, is highly predicted by the mass - velocity dispersion in the galactic bulge of low-mass, low-luminosity galaxies, as well as in a number of galaxies which contain active galactic nuclei \cite{Valluri:2005up, Safonova:2009td}. The evidence for SMBHs ranging from $10^6 M_\odot$ to $10^9M_\odot$, is quite abundant. They are thought to dwell at the centers of most galaxies. For various recent SMBHs mass estimation results and methods see for instance \cite{Beifiori:2011be, Li:2011ik, Rafiee:2011ry, Burkert:2010ki, Harris:2010zr}. The SNR in NGO is quite low for BH masses of less than about $5\times10^5 M_\odot$. Therefore, I take the lower end of the mass range to be $5\times10^6M_\odot$ and consider two other values of $25\times10^6M_\odot$ and a $10^8 M_\odot$ BH, to cover the interesting range of masses potentially observable in NGO. 
 
In the case of ET and aLIGO I have considered three IMBHs of mass 200$M_\odot$, 600$M_\odot$ and 1000$M_\odot$.
It is believed that BHs in this range are situated in the centers of many globular clusters. However, their existence is being questioned, the evidence is thought to be strong but circumstantial \cite{Miller:2008fi, Safonova:2009td, Noyola:2011pf, FregeauIMBH06, 1538-3881-135-1-182}. 
 
Concerning the mass ratios, only unequal mass binaries are presented, except for one case in aLIGO. We do not consider the equal mass case as such systems are not as likely to occur in nature as slightly asymmetric ones. For NGO we examine the mass ratios of 2, 10 and 20, whilst lower mass ratios of 2, 5 and 10 are considered for ET. We, therefore, have in total, 9 different sets of simulations for each detector. Let me emphasize here that there is a possibility that massive BH binaries in the early universe $(z \leq 10)$ will have a mass ratio significantly larger than one. For instance, in \cite{Volonteri:2004cf} it is suggested that low-redshift massive BH mergers occur predominantly with a mass ratio of 10 or higher.

\section{Signal detectability and Parameter estimation}
I will discuss the total signal to noise ratio, as well as the fractional errors in estimating the following 5 parameters: $\{M,\,j,\,q,\,D_{\rm L},\, \iota\}$. These errors are actually the quantities, $\sigma_\lambda=\sqrt{C_{\lambda \lambda}}$, which are computed from the covariance matrix, $C_{\ell m}$ \cite{Wainstein,Finn92,BalSatDhu96, SathyaSchutzLivRev09}. 

The results from all the distinct arrangements of the system (see Section \ref{subsec:simul}) are presented via cumulative frequency distribution plots. That is to say, the different system configurations are classified according to the value they render in the error of the observable quantity in question. The proportion of a number of occurrences in a small width of values should, to a good approximation, equal the probability that a randomly placed observer will measure that quantity to take this range of values. 

\vspace{-0.04in} 
\begin{figure*}
\begin{tabular}{|c|c|c|}
\hline
\includegraphics[width=0.31\textwidth]{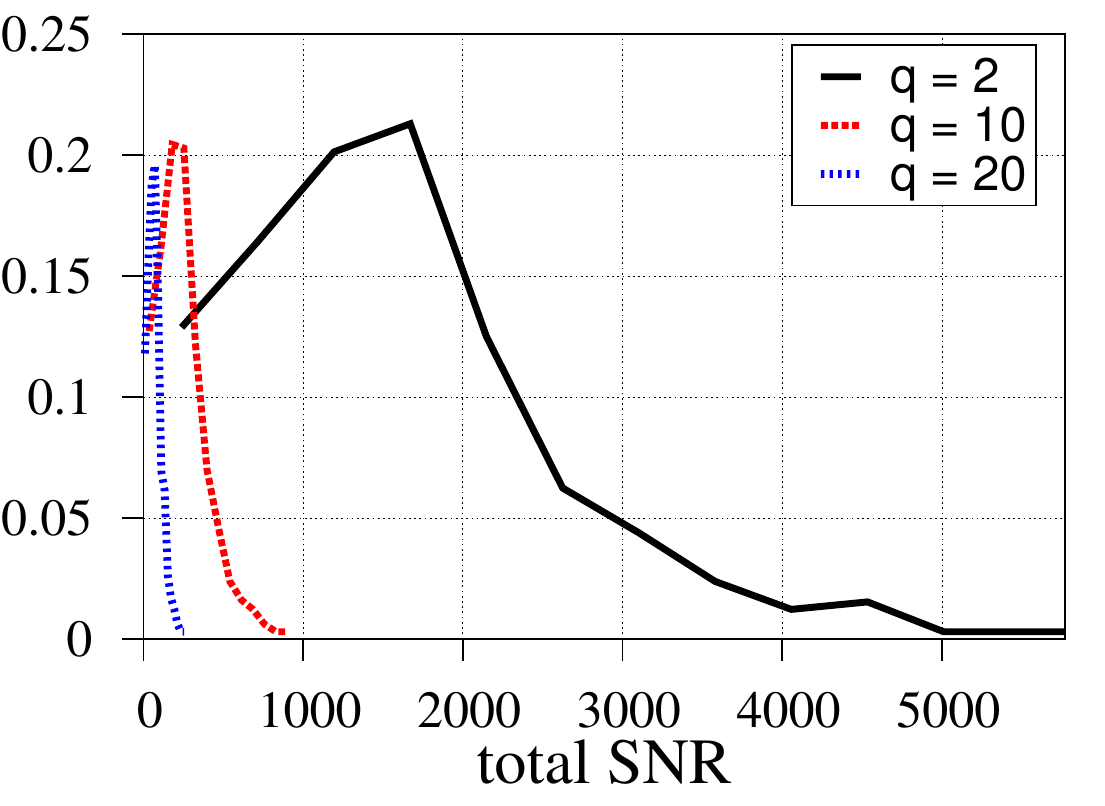}   &
\includegraphics[width=0.31\textwidth]{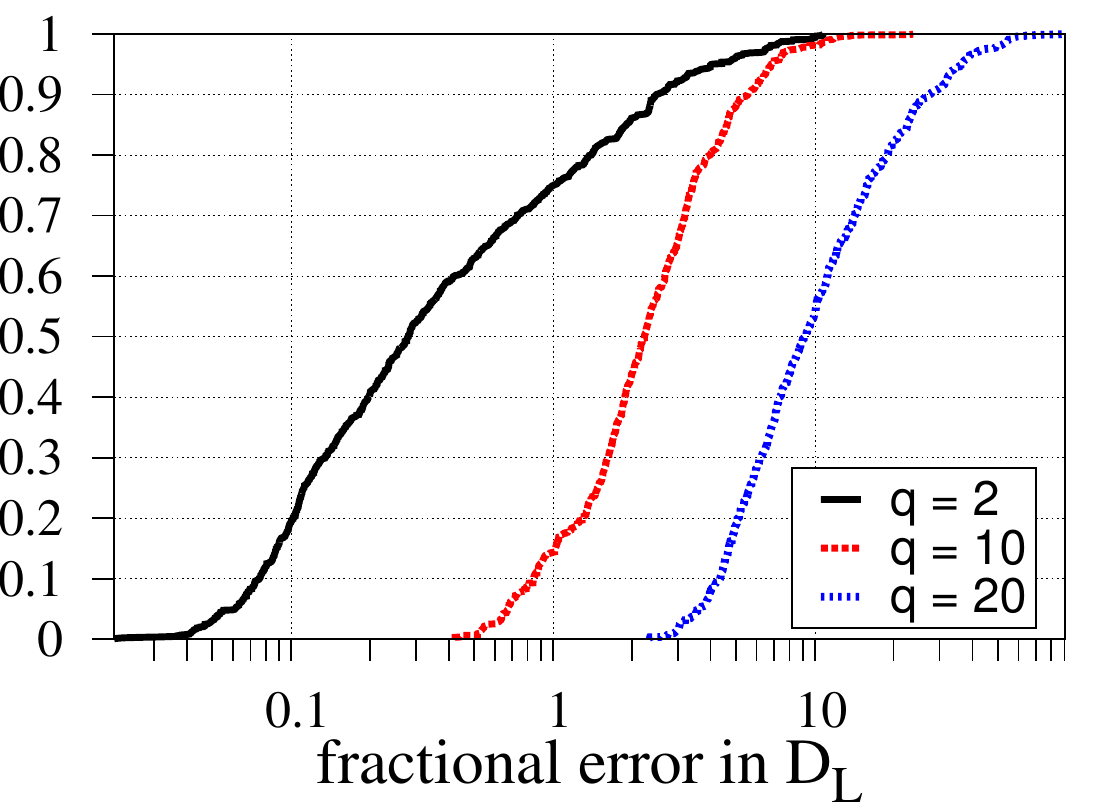}  &
\includegraphics[width=0.31\textwidth]{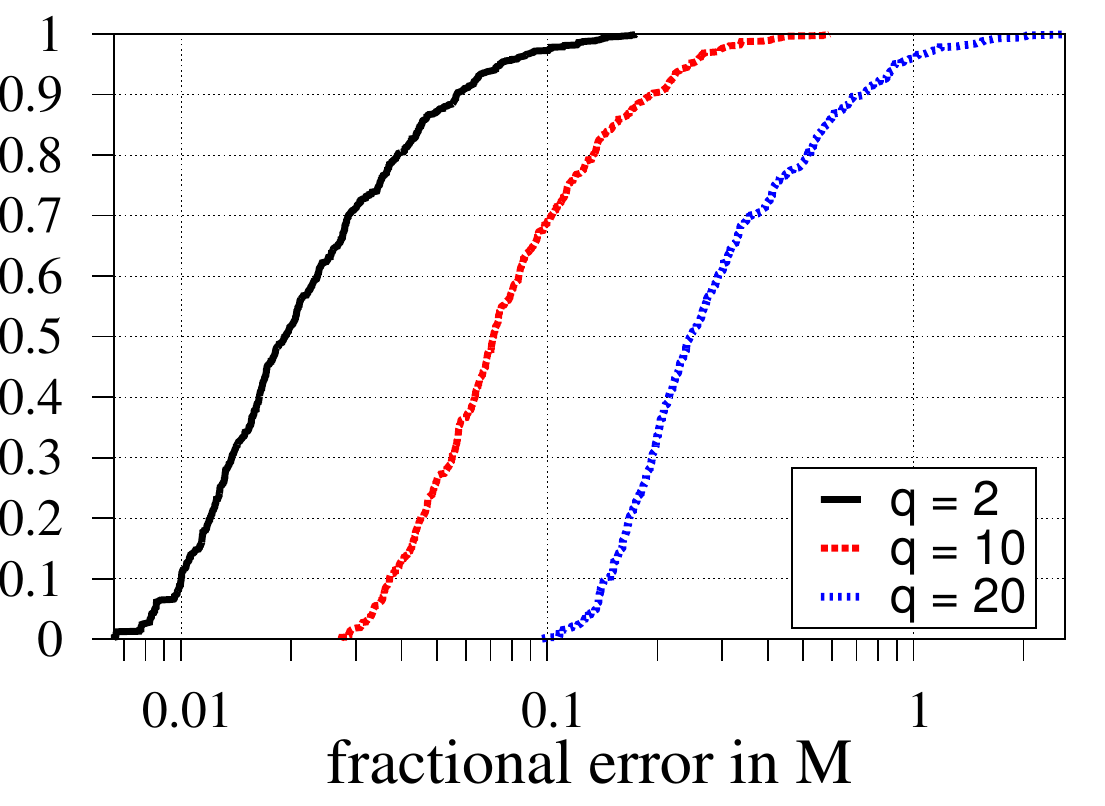}   \\
\hline
\includegraphics[width=0.31\textwidth]{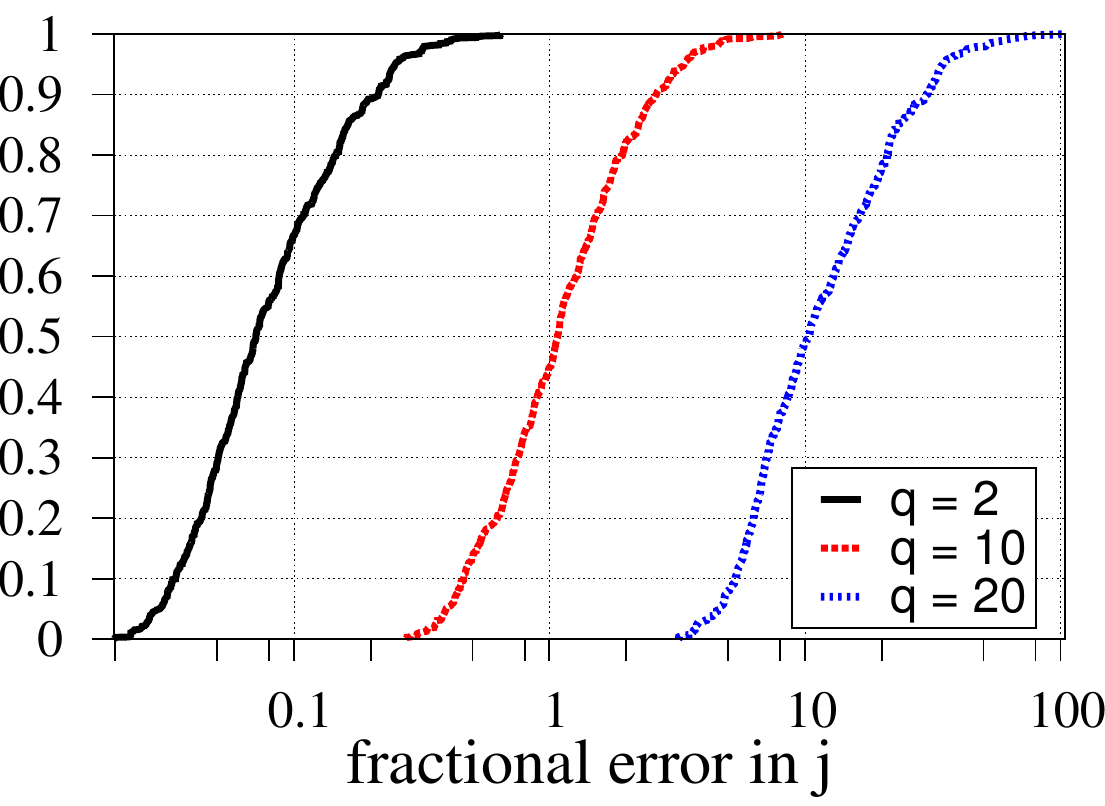}  &
\includegraphics[width=0.31\textwidth]{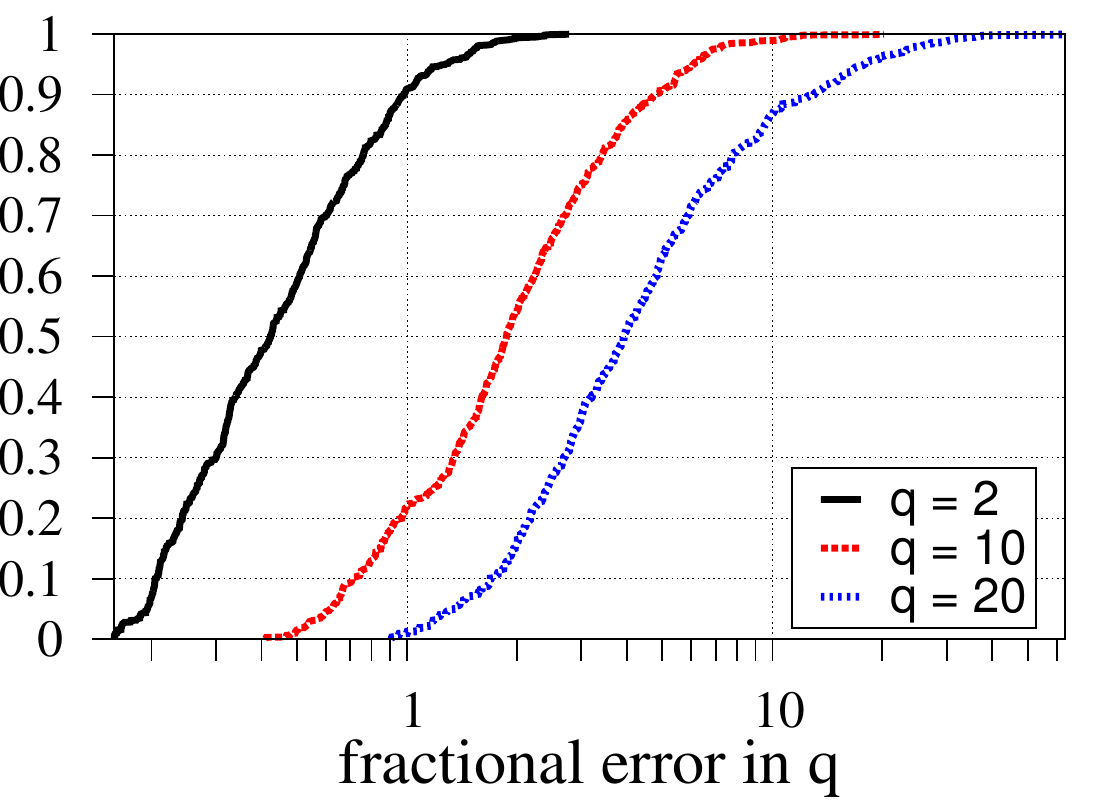} &
\includegraphics[width=0.31\textwidth]{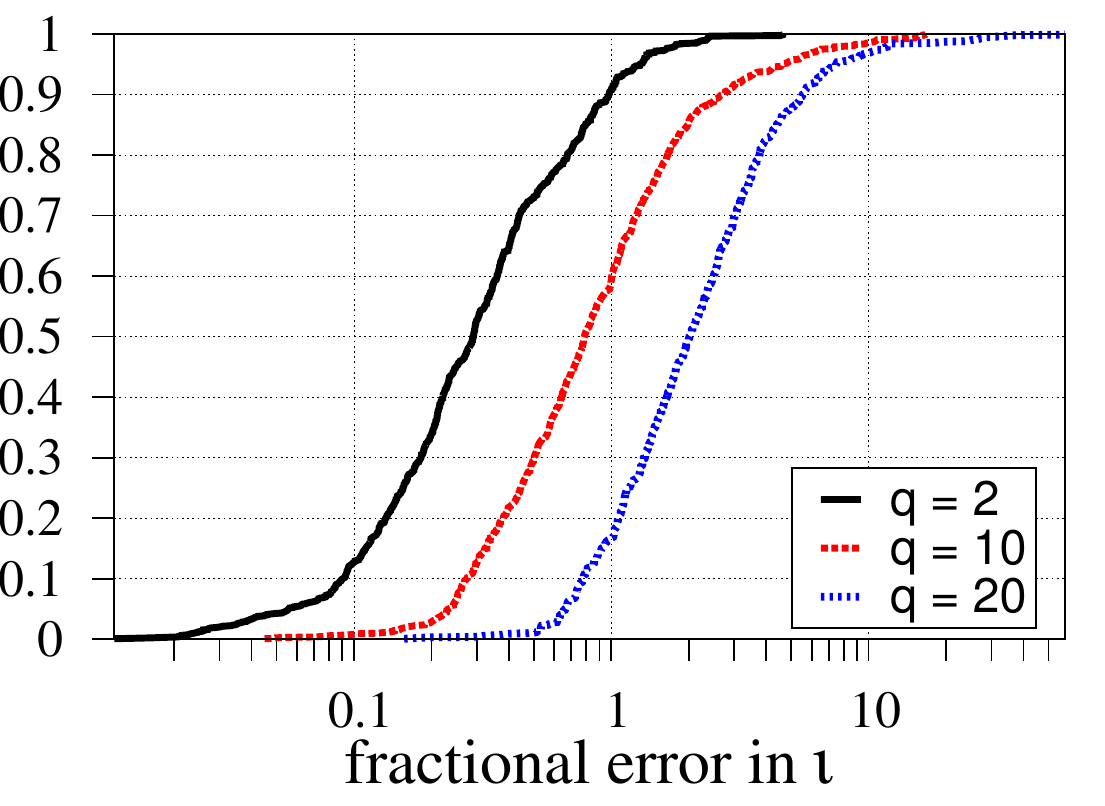}  \\
\hline
\end{tabular}
\caption{Same as in Figs. \ref{fig:5e6} and \ref{fig:25e6}, but for a $10^8M_{\odot}$ BH. The results slightly deteriorate with respect to the lower mass BH of Fig. \ref{fig:25e6}.}
\label{fig:1e8}
\end{figure*}
\vspace{-0.04in} 

\subsection{NGO}
The sensitivity curve\footnote{see https://lisa-light.aei.mpg.de/bin/view/DetectorConfigurations}
that we use is what is thought to be possible for NGO and it is contrasted with other LISA-like sensitivity curves in Fig. \ref{fig:PSDs}. It refers to a 4-link interferometer, comprised of one mother and two daughter spacecraft, having armlengths of $10^9\hspace{1pt}m$ and trailing a few degrees behind the Earth, in heliocentric orbit. 
  
The main noise contributions are the acceleration noise, the shot noise, as well as some other measurement noise. These are respectively:
\vspace{-0.08in} 
\begin{eqnarray}  
S_{acc, m}(f) &=& 1.37 \times 10^{-32} (1 + \frac{10^{-4}}{f}) f^{-4} \hspace{4pt} m^2 Hz^{-1}, \nonumber \\
S_{SN, m} &=& 5.25 \times 10^{-23} \hspace{4pt} m^2 Hz^{-1},  \nonumber \\
S_{OMN, m} &=& 6.28 \times 10^{-23} \hspace{4pt} m^2 Hz^{-1}. \nonumber
\end{eqnarray}
\vspace{-0.14in} 

The formula for the amplitude sensitivity curve is,
\vspace{-0.04in} 
\begin{equation}
\sqrt{S_{h}(f)} = \sqrt{5} \frac{2}{\sqrt{3}} T(f) \frac{\sqrt{4 S_{acc} + S_{SN} + S_{OMN}}}{L} \hspace{4pt} Hz^{-1/2},
\end{equation}
while the transfer function is
\vspace{-0.04in} 
\begin{equation}
T(f) = \sqrt{1 + \left(\frac{f}{0.41\left(\frac{c}{2L}\right)}\right)^2}. \nonumber
\end{equation}
with $L=10^9\hspace{1pt}m$ and $c=299,792,458$ metres per second. 
\vspace{-0.05in} 

 \begin{figure*}
\begin{tabular}{|c|c|c|}
\hline
\includegraphics[width=0.31\textwidth]{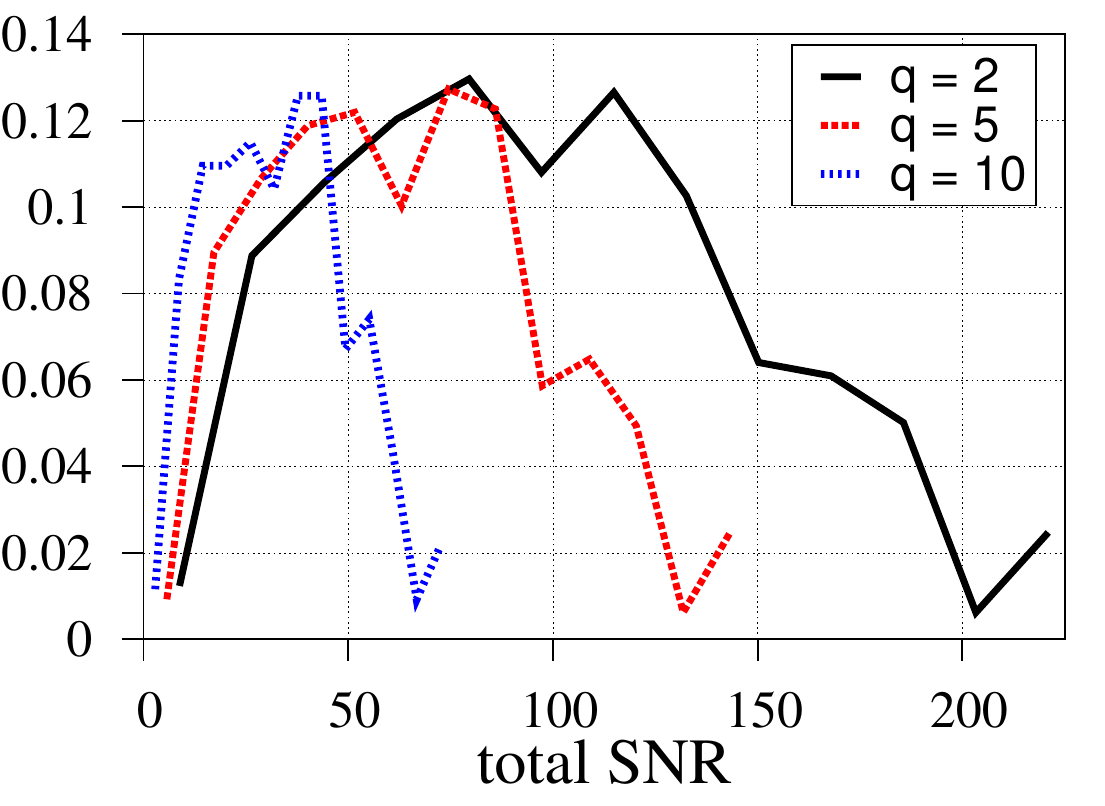}   &
\includegraphics[width=0.31\textwidth]{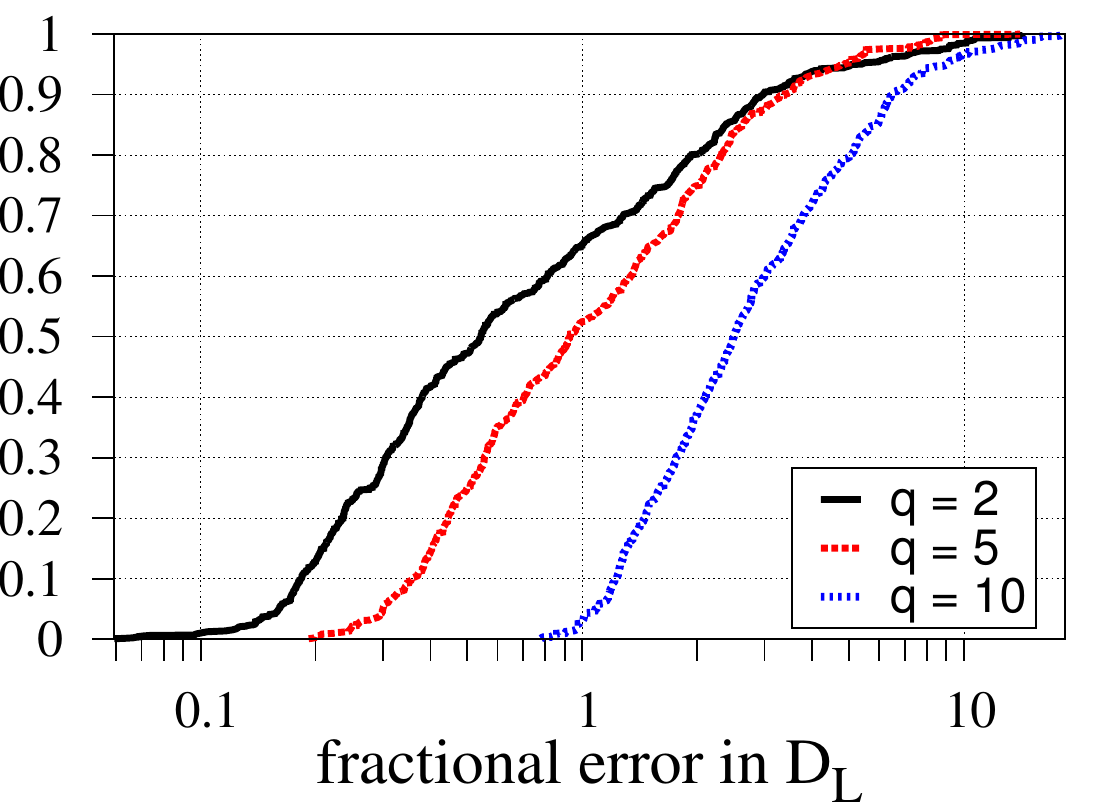}  &
\includegraphics[width=0.31\textwidth]{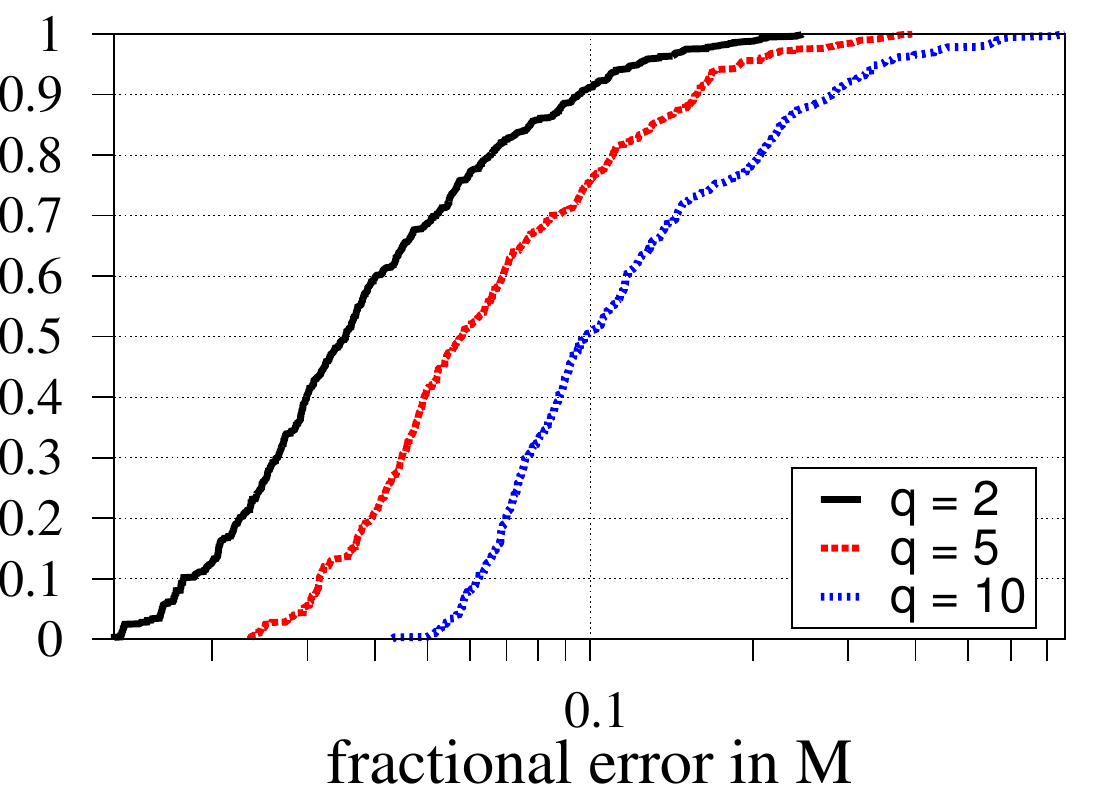}   \\
\hline
\includegraphics[width=0.31\textwidth]{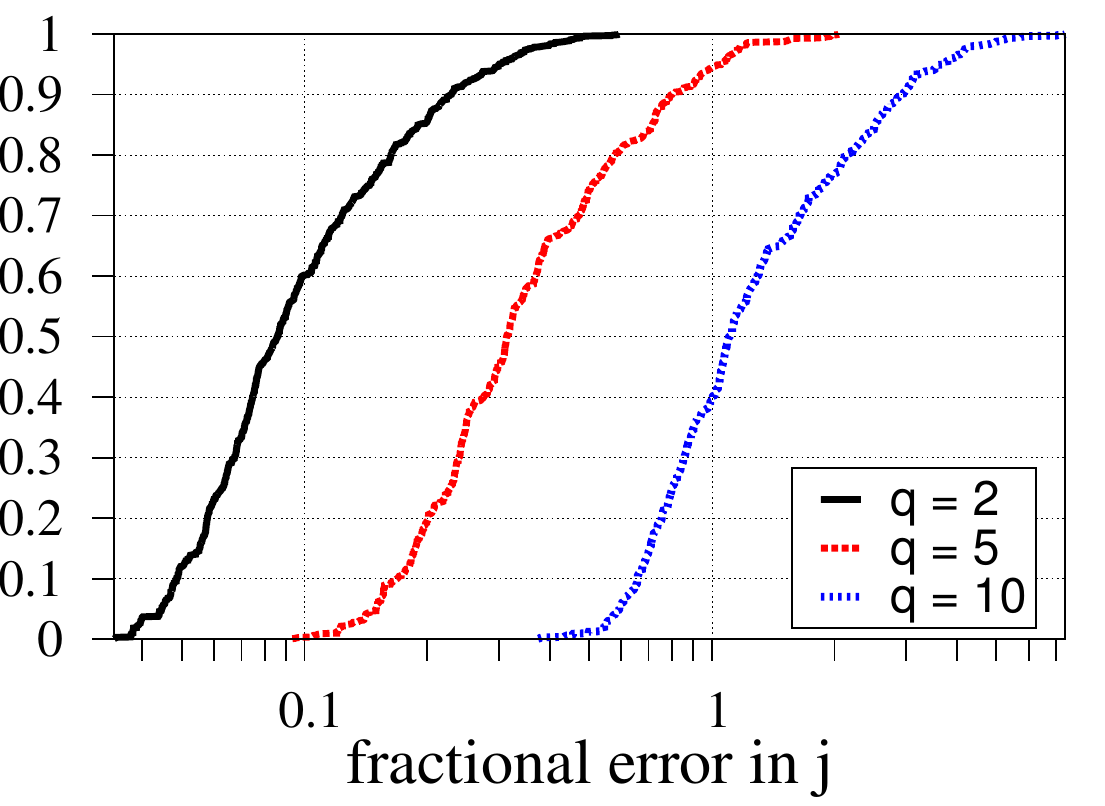}  &
\includegraphics[width=0.31\textwidth]{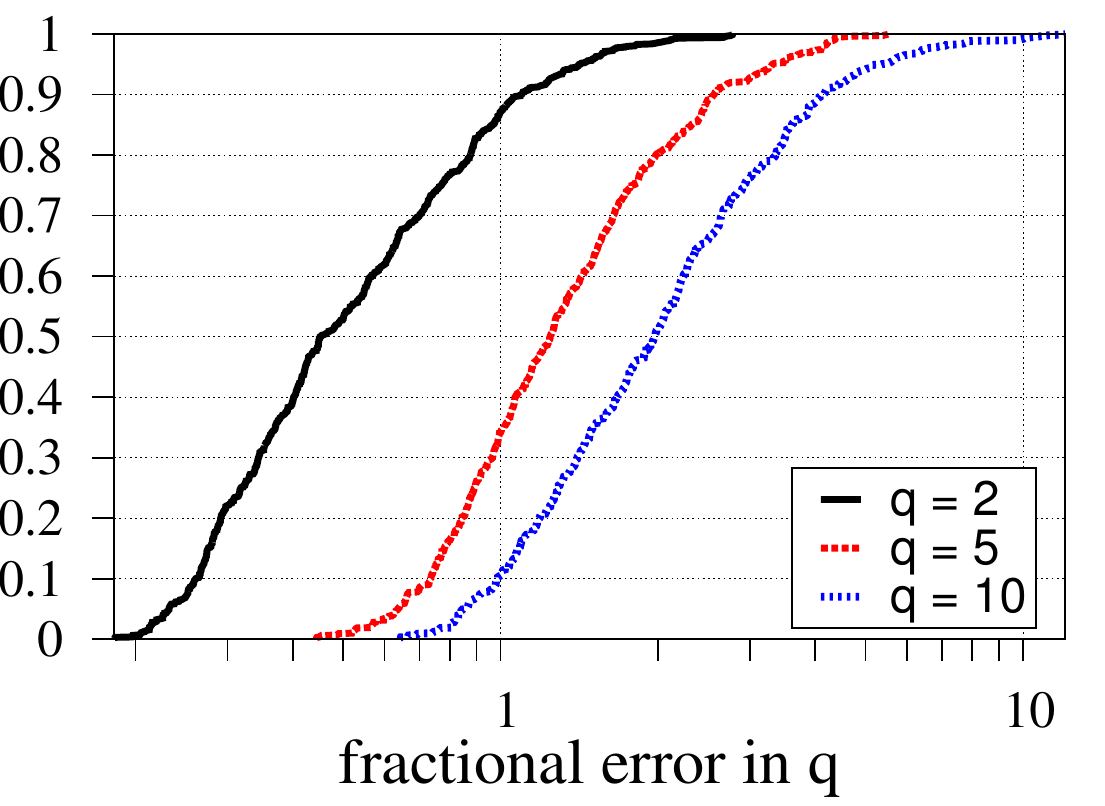} &
\includegraphics[width=0.31\textwidth]{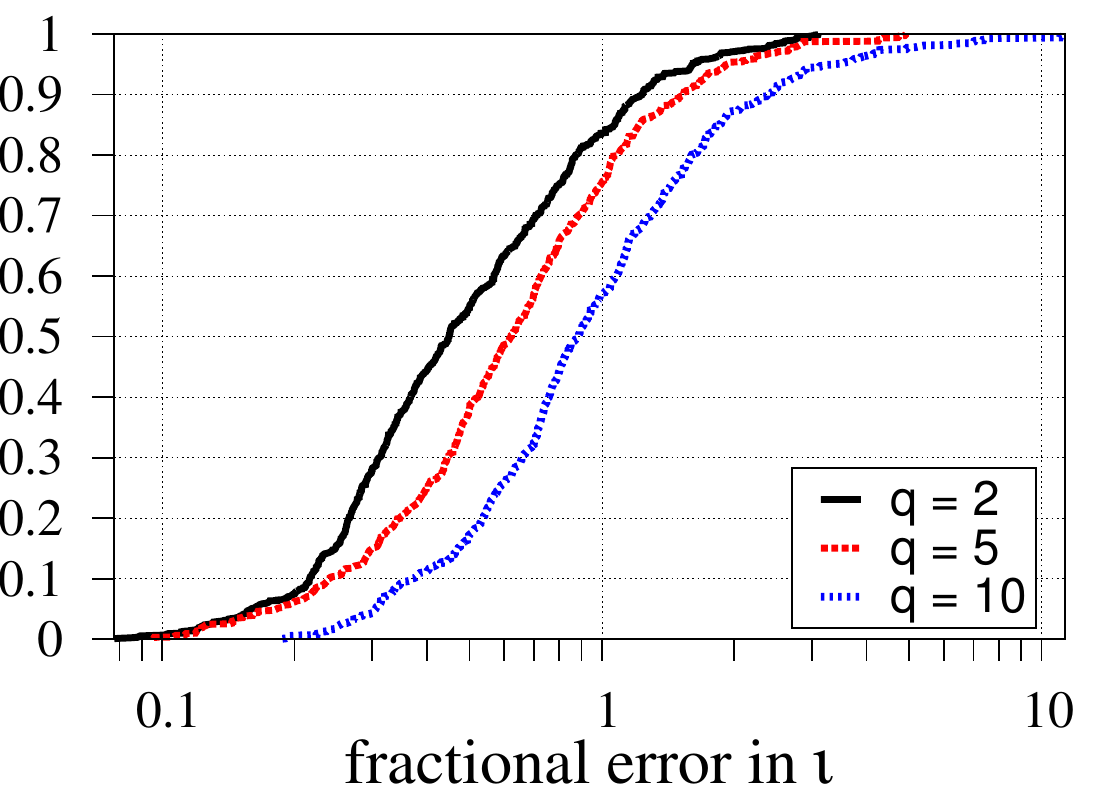}  \\
\hline
\end{tabular}
\caption{Similar graphs as those in Figs. \ref{fig:5e6} - \ref{fig:1e8}, although for the ET detector, concerning a IMBH of $200M_{\odot}$. The luminosity distance of the BH is now picked closer, at 1 Gpc ($z\simeq0.2$). Also, note that the progenitor mass ratios now take lower values, namely 2, 5 and 10.}
\label{fig:2e2}
\end{figure*}
\vspace{-0.04in} 

\subsection{Results for NGO}
Our results for NGO are plotted in Figs. \ref{fig:5e6}-\ref{fig:1e8}. 
For the $5\times10^6M_{\odot}$, $25\times10^6M_{\odot}$ and $10^8M_{\odot}$ BHs the SNR curves (top left subplots of Figs. \ref{fig:5e6}-\ref{fig:1e8}) peak at around 300, 600 and 1700 respectively. Relative frequency distributions were chosen for their plotting, as they portray clearly where the maximum occurs, as well as how they fall off. The higher mass ratio SNR curves have a similar outline. The fact that all of the curves resemble log-normal distributions, with steep risings and long tails, is mostly attributed to the fact that only a small fraction of the configurations, those close to the optimal orientation of the binary, will yield the highest SNRs.  

Regarding the rest of the plots in Figs. \ref{fig:5e6}-\ref{fig:1e8}, the general trend is that the mass and spin have comparable, low fractional errors, whilst the other group of parameters, namely the luminosity distance, mass ratio and inclination angle, yield an order of magnitude higher error values. This is not surprising, considering that the mass and spin have a direct effect on the modes' frequencies, $\omega_{\ell m n}$ and time-constants, $\tau_{\ell m n}$, quantities that determine to first order the shape of the ringdown waveform.  

\begin{figure*}
\begin{tabular}{|c|c|c|}
\hline
\includegraphics[width=0.31\textwidth]{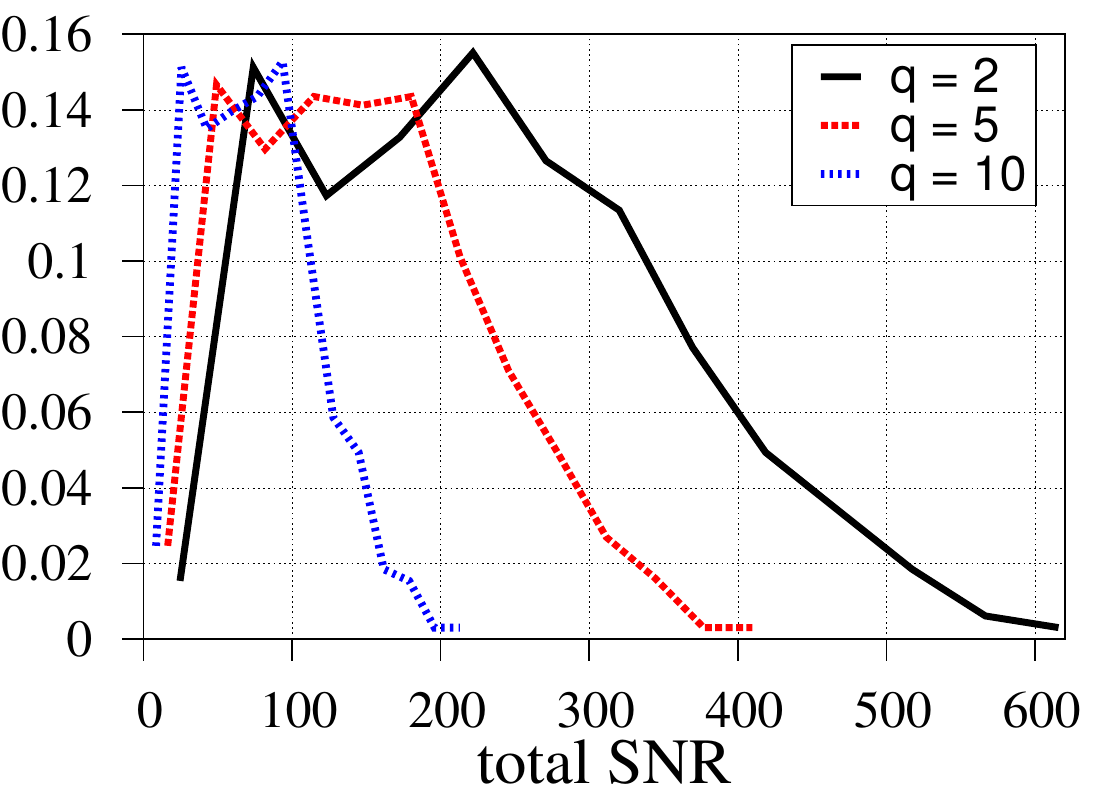}   &
\includegraphics[width=0.31\textwidth]{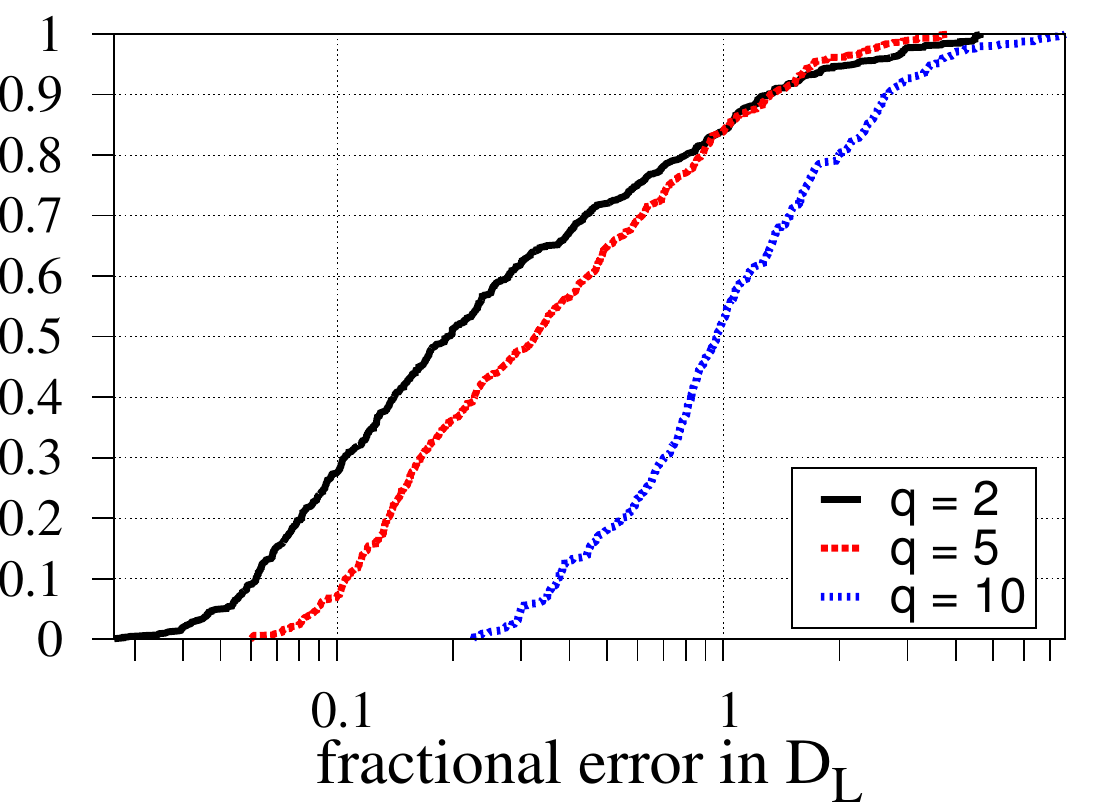}  &
\includegraphics[width=0.31\textwidth]{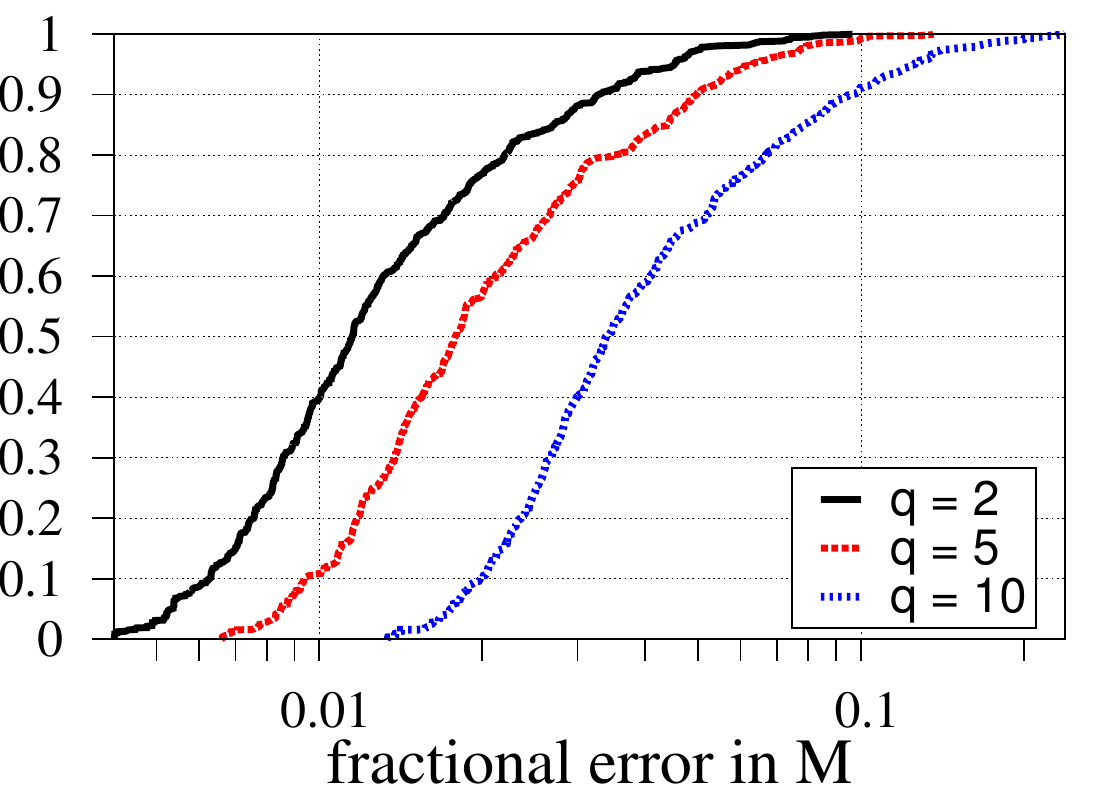}   \\
\hline
\includegraphics[width=0.31\textwidth]{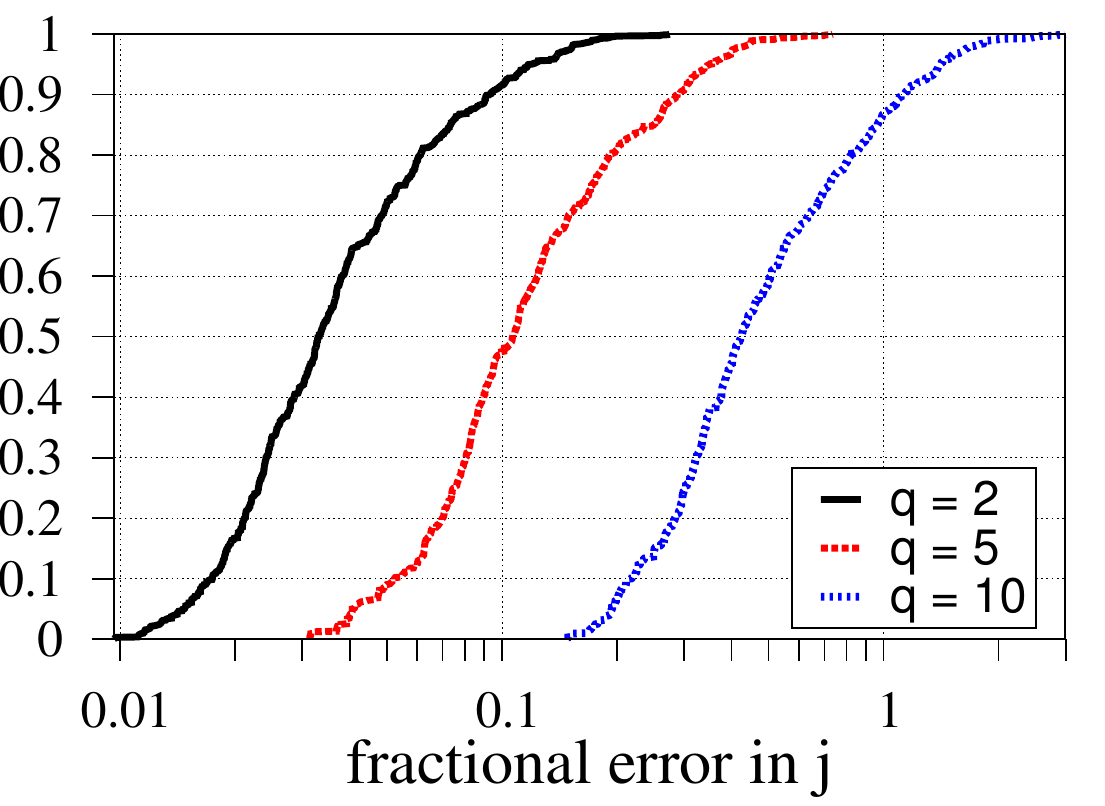}  &
\includegraphics[width=0.31\textwidth]{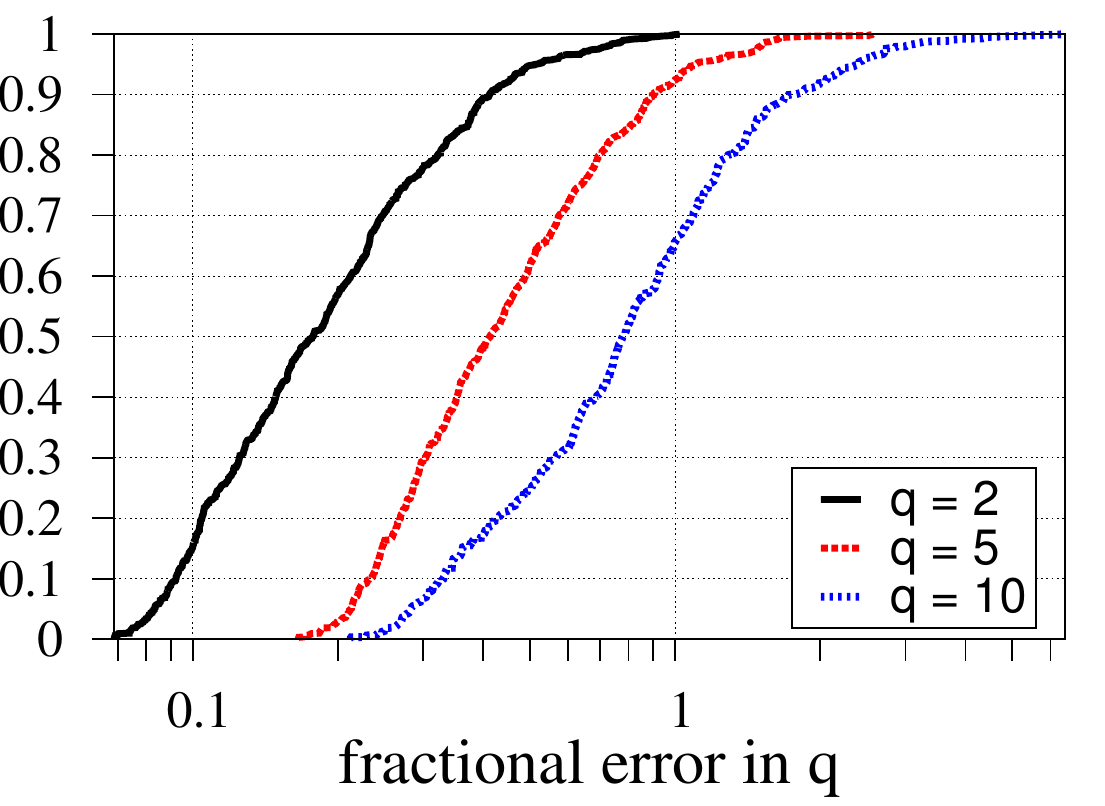} &
\includegraphics[width=0.31\textwidth]{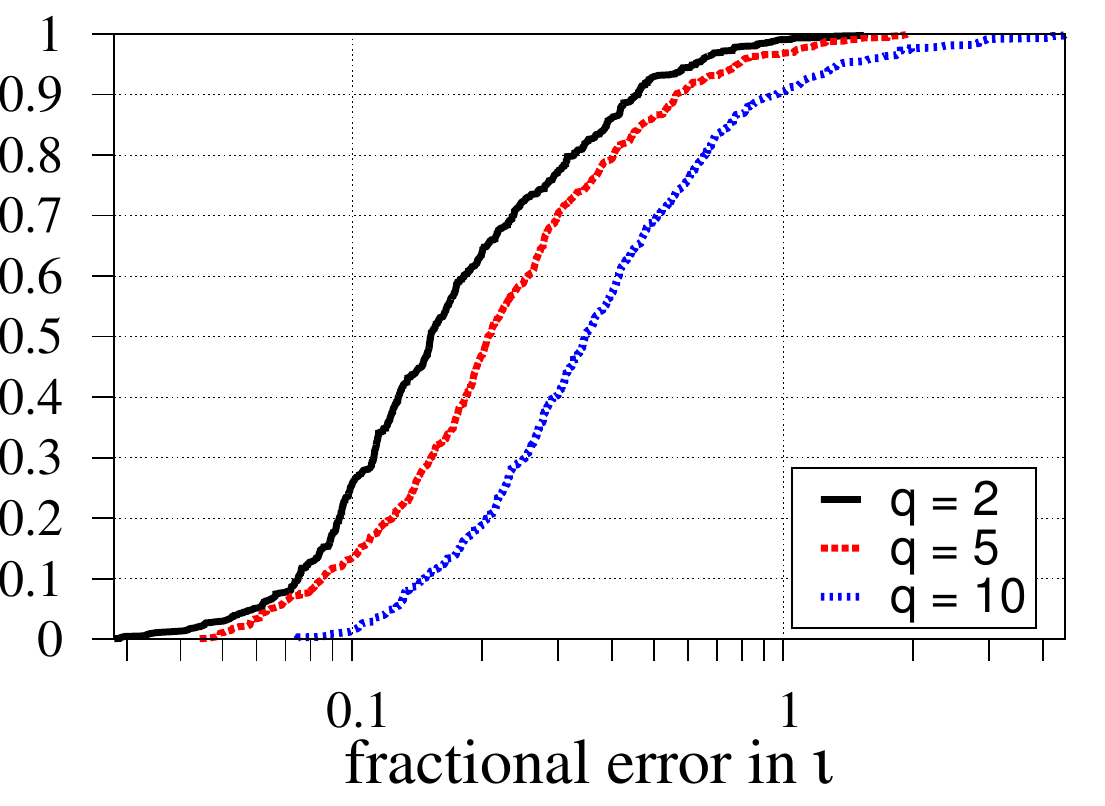}  \\
\hline
\end{tabular}
\caption{As in Fig. \ref{fig:2e2}, but for a $600M_\odot$ black hole.}
\label{fig:6e2}
\end{figure*}

Quoting a few numbers for a progenitor of mass ratio 2, the probability to get a binary merger event that will yield a mass error less than $1\%$ is correspondingly $58\%$, $55\%$ and $10\%$, as we go from the lower mass to the higher mass value. The lowest mass value actually gives the best results, whereas the signal's main power content takes place near the lowest sensitivity area of the NGO detector. The spin magnitude exhibits a similar trend with the probabilities to fall below a $10\%$ fractional error being $96\%$, $90\%$ and $67\%$ respectively. 

For the second group of parameters, I again quote how likely it is to do better than $10\%$. For the luminosity distance, the values are $38\%$, $41\%$ and $20\%$, corresponding to the BH masses $5\times10^6M_{\odot}$, $25\times10^6M_{\odot}$ and $10^8M_{\odot}$. The mass ratio is harder to determine accurately, with the likelihoods being $32\%$, $15\%$ and only $2\%$ respectively. Lastly, there is a $40\%$, $37\%$ and $13\%$ likelihood of achieving an accuracy better than $10\%$ in the inclination angle.
\label{subsub:NGO}    

\subsection{Einstein Telescope and Advanced LIGO}
ET's very low sensitivity curve accounts for impressive results in the mass range $\sim500M_\odot$ to $\sim1000M_\odot$ and for mass ratios between 1 and 5. I consider the sensitivity curve designated ET-B \cite{Hild:2009ns}, whose noise power spectral density is given by $S_h(f) = 10^{-50}h_n(f)^2\,{\rm Hz}^{-1}$, with: 
\vspace{-0.05in} 
\begin{eqnarray}
h_n(f) & = & 2.39\times 10^{-27}\, x^{-15.64} + 0.349\, x^{-2.145} \nonumber \\ 
       & + & 1.76\, x^{-0.12} + 0.409\, x^{1.10},
\end{eqnarray}
where $x = f/100\,{\rm Hz}.$ 

As for advanced LIGO, the noise spectral density is \footnote{This fit was provided by C. Capano, Syracuse University and is tuned for detecting binary neutron stars.}
\vspace{-0.05in} 
\begin{eqnarray}
S_h(f) & = & 10^{-49} \Biggl [ 10^{16-4\,(f-7.9)^2} 
+ 0.08\, x^{-4.69} \nonumber \\ & + & 123.35\, 
\frac{1-0.23\, x^2+0.0764\,x^4}{1+0.17\, x^2}\Biggr ]\,{\rm Hz}^{-1},
\end{eqnarray}
where $x=f/215\,{\rm Hz}.$ 
\vspace{-0.03in} 

\subsection{Results for ET and aLIGO}
\begin{figure*}
\begin{tabular}{|c|c|c|}
\hline
\includegraphics[width=0.31\textwidth]{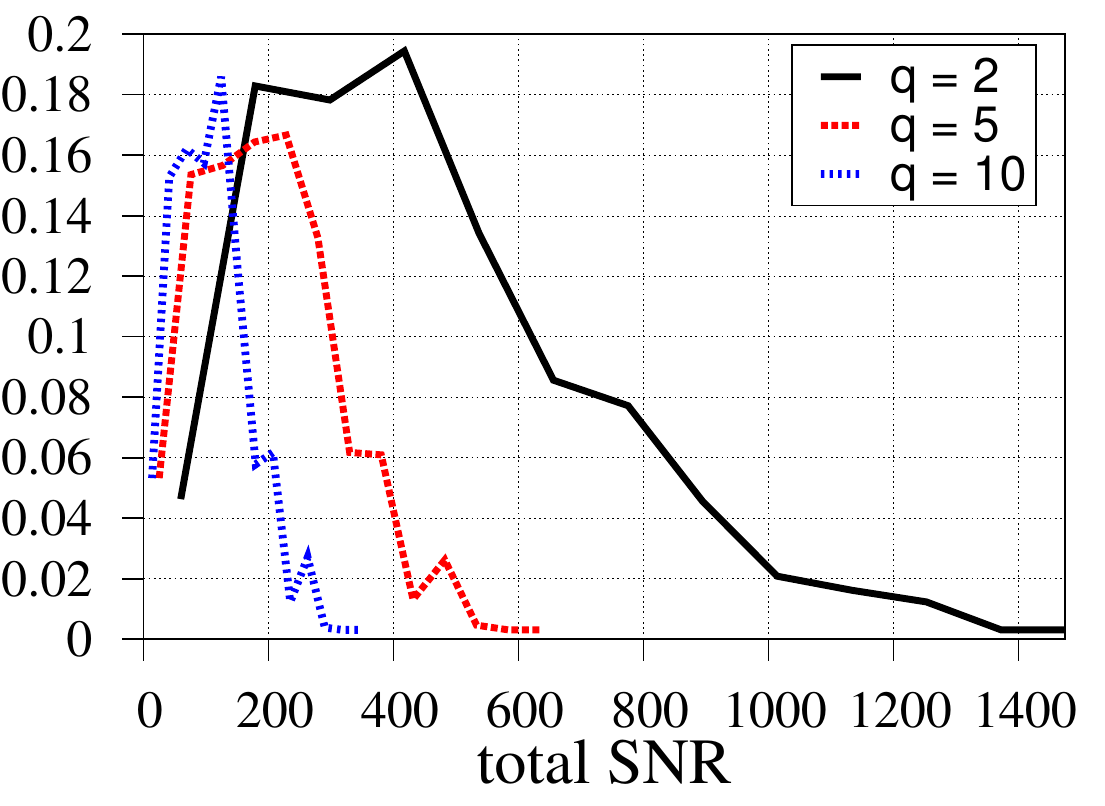}   &
\includegraphics[width=0.31\textwidth]{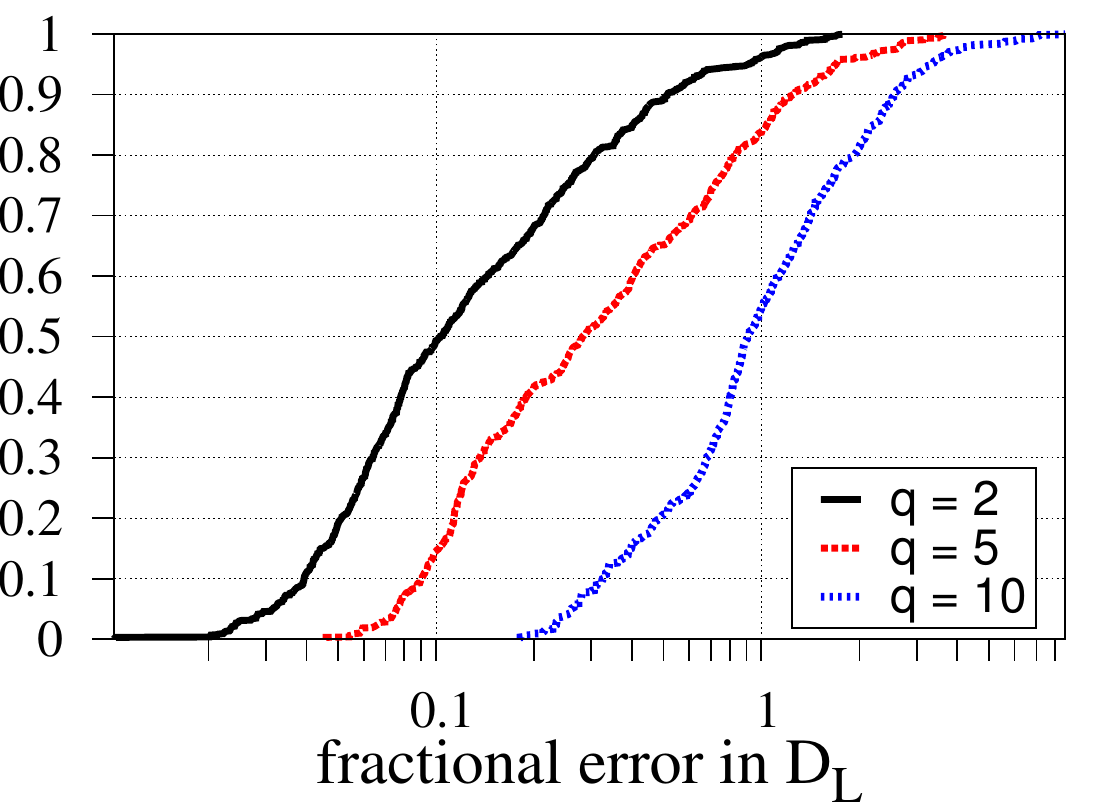}  &
\includegraphics[width=0.31\textwidth]{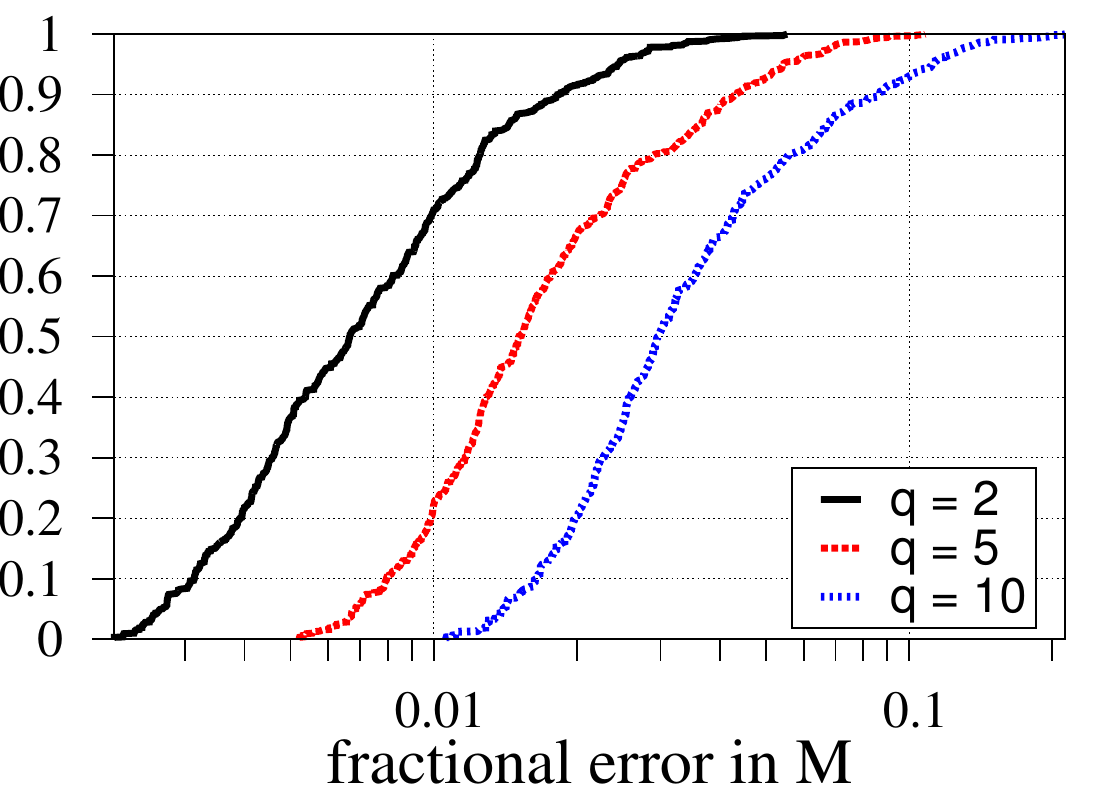}   \\
\hline
\includegraphics[width=0.31\textwidth]{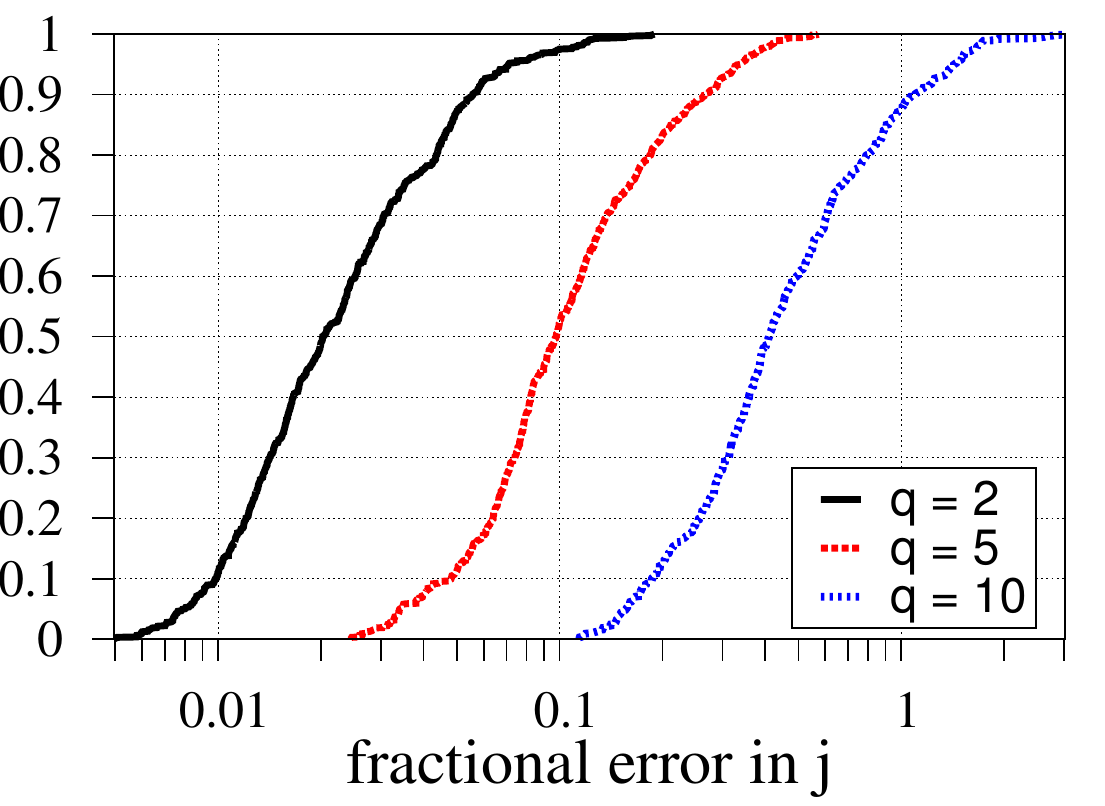}  &
\includegraphics[width=0.31\textwidth]{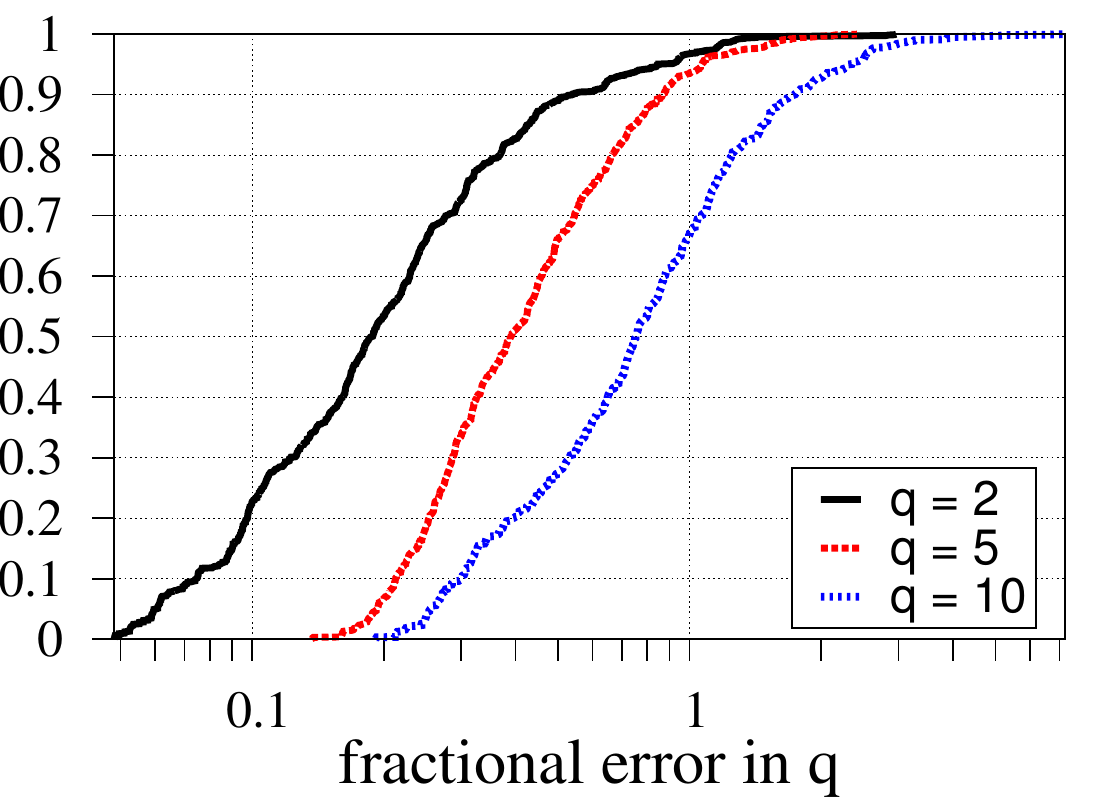} &
\includegraphics[width=0.31\textwidth]{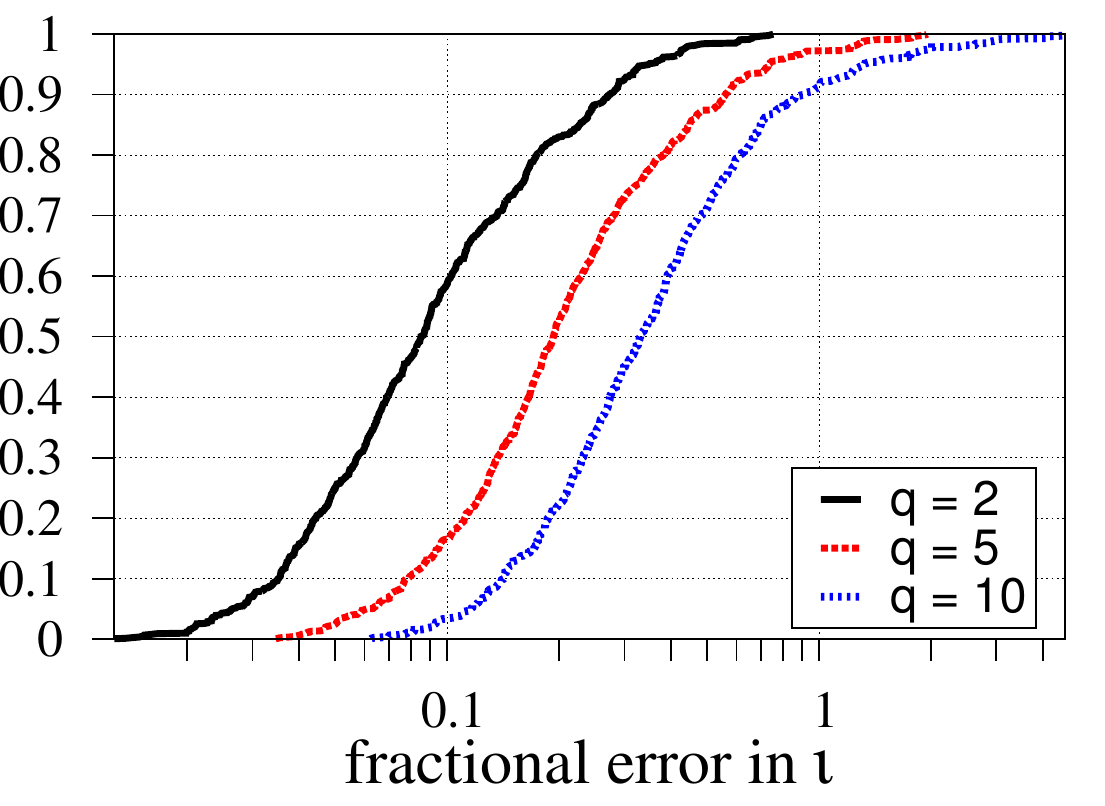}  \\
\hline
\end{tabular}
\caption{As in Figs. \ref{fig:2e2} and \ref{fig:6e2}, but for a $1000M_{\odot}$ BH, a sweet spot in ET. The results are very encouraging for this size of BH and distance, with a very high probability to get errors lower than $10\%$ in most of the parameters at mass ratios 2-5.}
\label{fig:1e3}
\end{figure*}

\begin{figure*}
\begin{tabular}{|c|c|c|}
\hline
\includegraphics[width=0.31\textwidth]{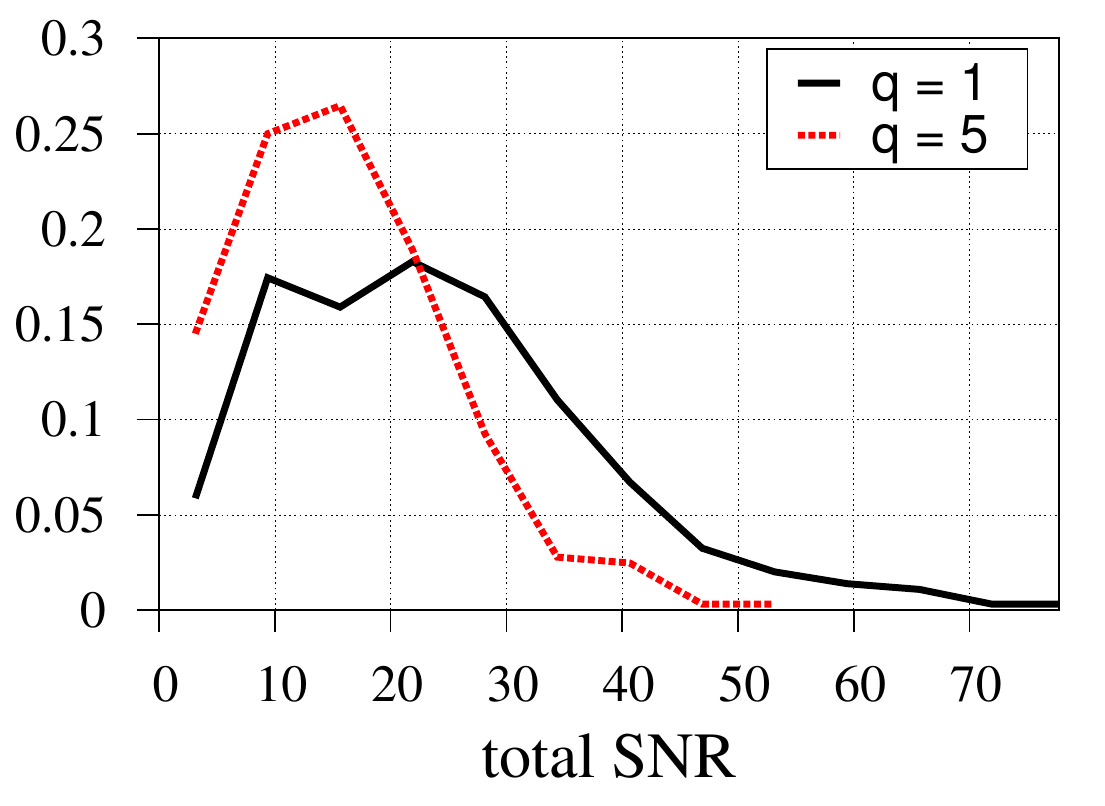}   &
\includegraphics[width=0.31\textwidth]{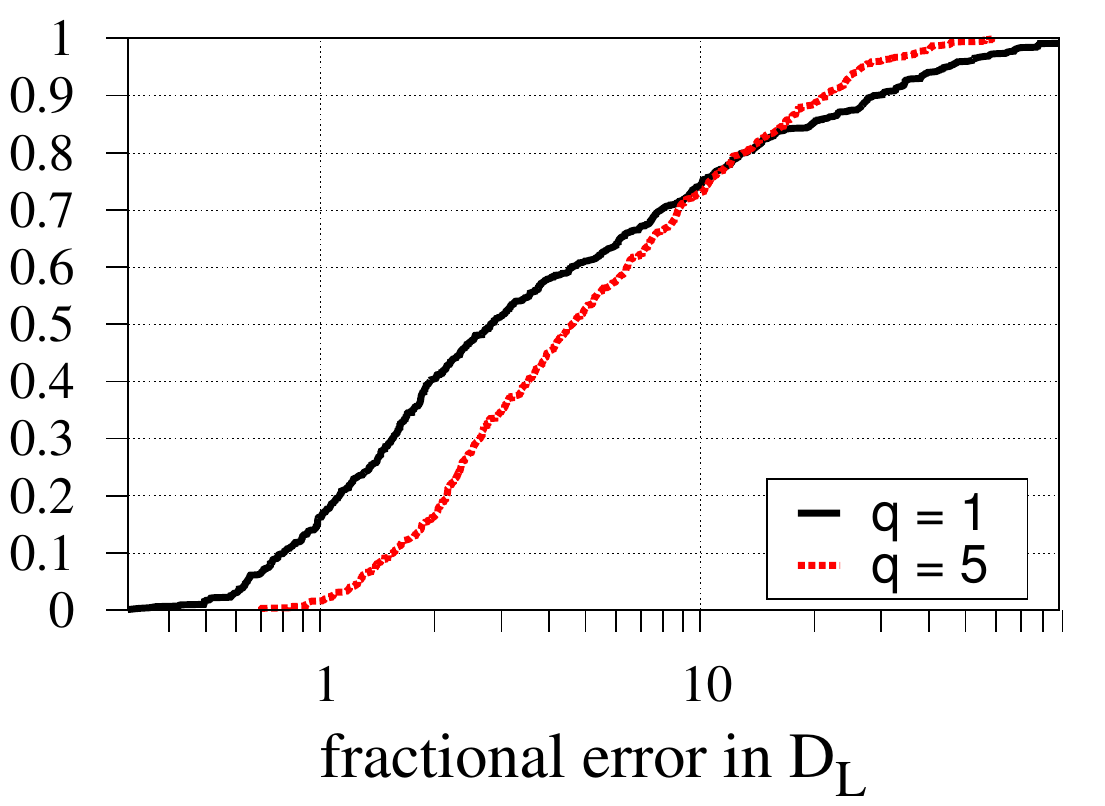}  &
\includegraphics[width=0.31\textwidth]{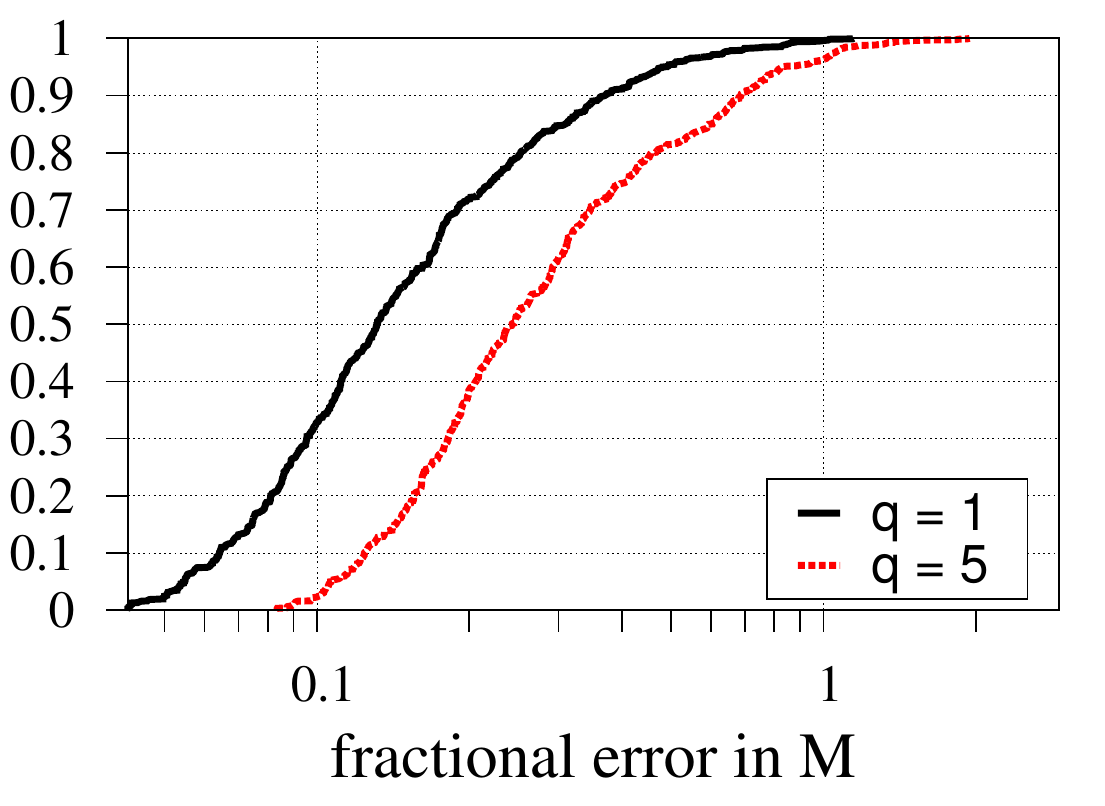}   \\
\hline
\includegraphics[width=0.31\textwidth]{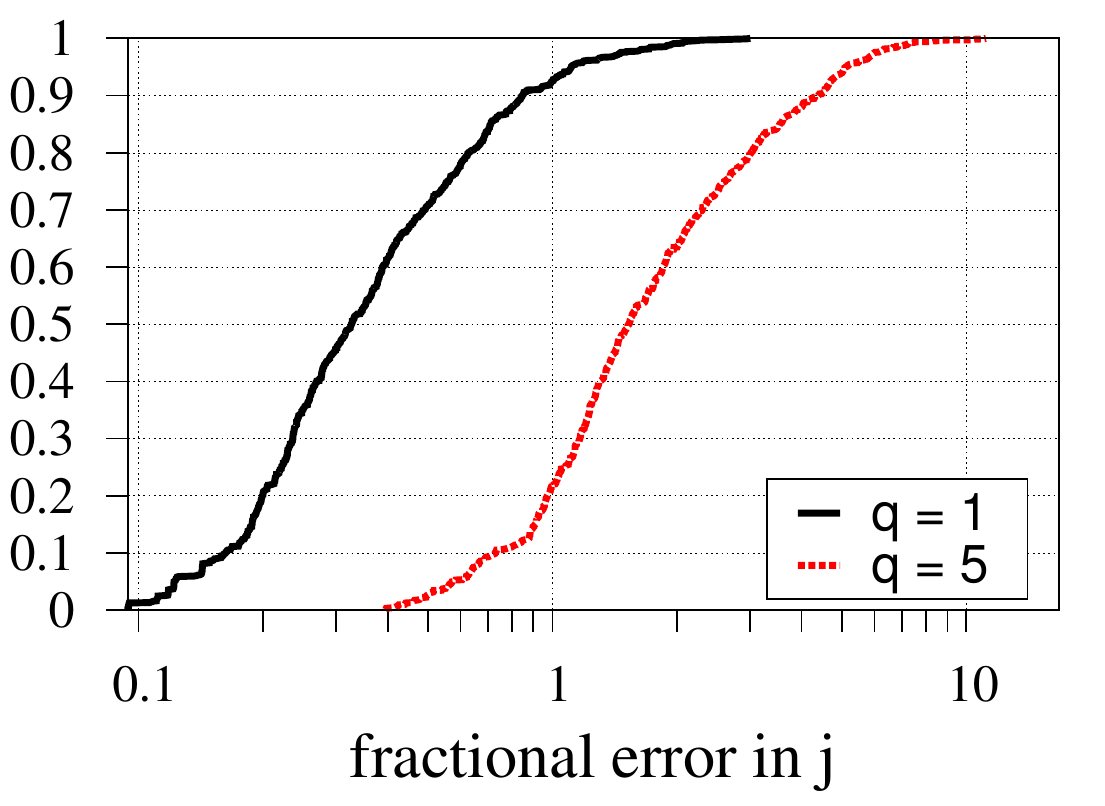}  &
\includegraphics[width=0.31\textwidth]{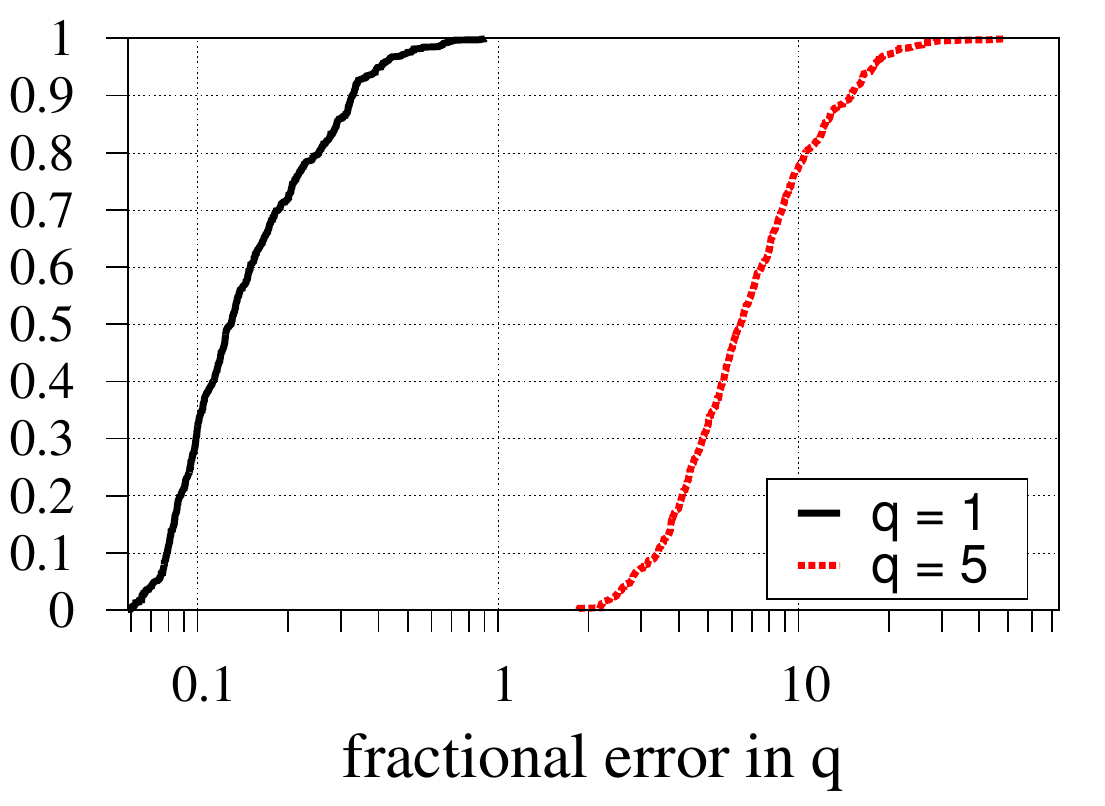} &
\includegraphics[width=0.31\textwidth]{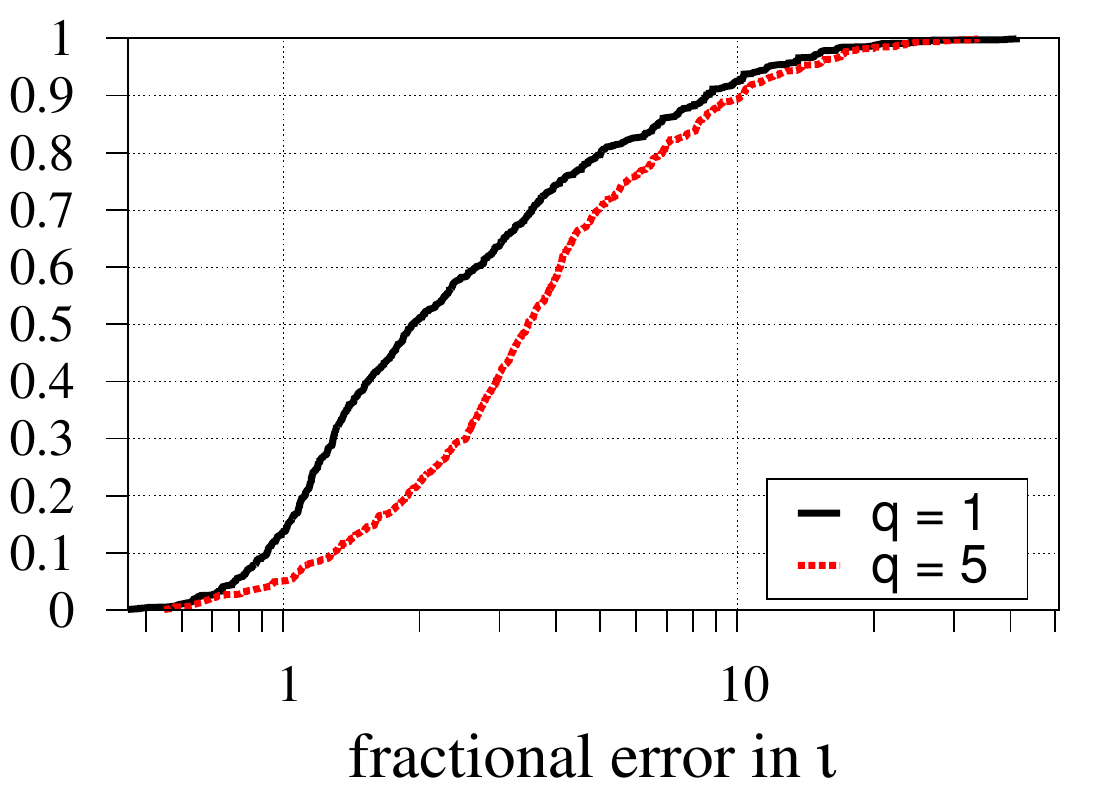}  \\
\hline
\end{tabular}
\caption{Frequency distributions involving the most optimistic scenario for advanced LIGO: a $1000M_\odot$ BH which is the merger of an equal mass binary, that is $q=1$. The BH is again situated at 1 Gpc. A mass ratio of 5 is shown as well for contrasting.}
\label{fig:1e3aLIGO}
\end{figure*}

The results obtained for ET and advanced
LIGO are plotted in Figs. \ref{fig:2e2}-\ref{fig:1e3aLIGO}. We fix the luminosity distance of the BH to be 1 Gpc. 
For the lowest mass considered (a $200M_\odot$ BH) although the SNR could be pretty high (in the range 30-200), errors in the estimation of
parameters are poor (see Fig. \ref{fig:2e2}). $\{D_L, q, \iota\}$ have $50\%$ probability to be measured to an accuracy of $\sim50\%$, while errors on $\{M, j\}$ are $90\%$ and $60\%$ likely to be below $10\%$. For the higher mass ratios the results, as expected, are worse.

The results for $600M_\odot$ and $1000M_\odot$ BHs are shown in Figs. \ref{fig:6e2} and \ref{fig:1e3}. The parameter estimation accuracies for these systems observed with ET is almost as good as that for a SMBH with NGO. Referring to the heaviest BH and mass ratio in the range 2-5, it is $100\%$-$99\%$ and $96\%$-$53\%$ likely to acquire errors below $10\%$ for the BH mass and spin. For the $\{D_L, q, \iota\}$, the efficiencies are correspondingly $50\%$-$15\%$, $23\%$-$0\%$ and $60\%$-$17\%$. The results are similar for the $600M_\odot$ case (see Fig. \ref{fig:6e2}).
 
Fig. \ref{fig:1e3aLIGO} shows two examples for aLIGO: a BH resulting from an equal mass binary and one for which the mass ratio is 5. Most configurations give ringdown SNR values in the range 10 to 30, not large enough for a good estimation of parameters. This translates to a $30\%$ likelihood for a fair measurement of $10\%$ accuracy in the mass and mass ratio, while the luminosity distance, spin and inclination angles are all measured to accuracies far worse than $10\%$. 
 \vspace{-0.04in} 

\vspace{-0.05in} 
\vspace{-0.08in} 

\section{Astrophysical implications}
\subsection{Supermassive black holes}
The presence of SMBHs in the centers of massive galaxies seems to be a well established fact. Detecting their 
gravitational wave signals can give additional clues on their spatial\footnote{Unless the inspiral phase is inside NGO's band, it will not be possible to determine the sky position of the signal. The only hope in this case would be the existence of an electromagnetic counterpart to the merger event.} and mass distribution, as well as help discriminate among the different scenarios of their formation and growth. This could be done, for instance, by measuring the BH mass function as a function of red-shift. Additionally, determining the mass ratios \cite{PhysRevD.85.024018} of these early universe merger events will be an important piece of information in selecting out the current models on SMBH formation. Note that, in NGO, BHs of mass higher than $10^{7-8}M_\odot$ are visible almost entirely due to the ringdown signal that they emit rather than the inspiral signal.

Several studies have been realized in the field of predicting the coalescence rates of SMBHs \cite{Filloux:2011eq, Berti06, 0264-9381-20-10-304, Rhook:2005pt, Erickcek:2006xc, Micic:2007vd, Wen:2011xc, deFreitasPacheco:2006bh}. Let us admit an event rate of $\sim10$ $yr^{-1}$ at $z\simeq1$, which ascertains a scenario of most efficient BH coalescence \cite{Menou:2001hb}. Then, assuming that the BH masses are both around $5\times10^6M_\odot$, Figs. \ref{fig:5e6} and \ref{fig:25e6}, there is a good chance that in 6 of the events, the BH mass will be measured more accurately than $1\%$ and that in 9 of the events the BH spin magnitude will have an accuracy better than $10\%$. As for the parameters $\{D_L, q, \iota\}$, in approximately 3-4 of the events they will feature errors lower than $10\%$, while in 7 of the events, the errors will lie below $30\%$ for $D_L$ and below $20\%$ for $q$ and $\iota$. 

Let me clarify that ringdown signals in NGO, will be detectable out to about $z\simeq5$, but the distance of $z=1$ was chosen to make the quote on the errors. If we take the distance at $z=3$, an optimistic rate of merger events will be of the order of 100 \cite{Menou:2001hb}. If we assume that they all involve BHs of masses $\simeq2$-$3\times10^6M_\odot$, then in around 90 of them the error\footnote{An additional simulation at $z=3$, of a $5\times10^6M_\odot$ BH, with a progenitor mass ratio of 2 was performed.} in the mass will be below $10\%$ and in 60 of them the spin magnitude error will fall below the $10\%$ threshold. For the rest of the parameters, about 50 of the observed events should yield measurements better than $50\%$ accuracy. Although this looks poor at the outset, it should suffice for a statistical test of different models of BH formation and growth.
 \vspace{-0.04in} 

\vspace{-0.15in} 

\subsection{Intermediate-mass black holes}
The existence of intermediate-mass black holes remains uncertain, as is their mass distribution. Colliding globular clusters in interacting galaxies could be a mechanism to obtain a compact binary IMBH system \cite{AmaroSeoane:2006py, AmaroSeoane:2009ui}. Another possibility could be the formation of a binary IMBH inside a young dense stellar cluster, especially when the fraction of binary stars is adequately high \cite{FregeauIMBH06}.

Estimates of IMBH-IMBH coalescence rates can be found in \cite{rates:2010cf, FregeauIMBH06, AmaroSeoane:2009ui, Mandel:2009dn}. A relatively optimistic rate is $R_{opt} =$ 0.007 $GC^{-1} Gyr^{-1}$, where it has been assumed that $10\%$ of star clusters are sufficiently massive and have a sufficient stellar binary fraction to form an IMBH-IMBH binary once in their lifetime, taken at 13.8 Gyr. The maximum possible rate would come from assuming that all of the star clusters satisfied the above conditions. The corresponding rate value is then $R_{max} =$ 0.07 $GC^{-1} Gyr^{-1}$.

If the number of relatively luminous galaxies within a distance of 1 Gpc is approximately $5.3\times 10^7$, \cite{White:2011qf} and the number of young dense stellar clusters per such galaxy is of the order of 100, then the optimistic estimate gives 0.037 events per year, while the maximum rate would be 0.37 events per year. An event within this distance will in some likelihood, involve IMBHs with masses between $\sim6\times10^2M_\odot$ and $\sim10^3 M_\odot$ which means a relatively fair chance for errors to be low for several of the
parameters using Einstein Telescope (see Figs. \ref{fig:6e2}, \ref{fig:1e3}).

\section{Conclusions}
The present study constitutes a sensible and realistic approach to the subject of parameter estimation from a multimodal ringdown signal, inasmuch as it is supposed to be emitted from a merged binary in a generic configuration. Parameters such as the inclination angle $\iota$, along with the sky location $\theta$ and $\varphi$ and the signal polarization $\psi$ have an effect on the observed quasi-normal mode spectrum. Their impact on the detectability and parameter estimation has been assessed, by performing a large sample of Fisher-matrix analysis simulations, allowing for a simple statistical analysis of the results. 

I am quoting frequency distributions for the errors at the representative distances of $z\simeq1$ (6.73 Gpc) for supermassive BHs, in NGO, and of $z\simeq0.2$ (1 Gpc) for intermediate-mass BHs, in the Einstein Telescope and advanced LIGO. An additional simulation was performed for a $5\times10^6M_\odot$ BH at $z\simeq3$. The results are quite satisfactory in determining the mass and spin, especially for the low mass ratios from 1 to $\sim5$, where in typically $90\%$ of the cases their errors fall below $10\%$. The effectiveness in measuring the luminosity distance, mass ratio and inclination angle is almost an order of magnitude worse.  

The effects of these results on, as much as possible, realistic event rates in NGO and ET are discussed. The likelihood to have a waveform parameter measured to an accuracy of a certain threshold translates to the same proportion of observed events featuring error values below that threshold. As an example, if supermassive BHs within a luminosity distance of $z=1$ coalesce at a rate of $\simeq10$, then NGO could act as a SMBH dynamics probe, as almost all of these events will yield very low errors $1\%-10\%$ in the BH mass and spin, while in half of the events the errors in the luminosity distance, mass ratio and inclination angle will be of the order of $20\%$.    

\section{Acknowledgements}
I am grateful to B.S. Sathyaprakash for his supervision and exceptional advice throughout this study. I would like to thank Mark Hannam for discussions and him and Sascha Husa for making available numerical relativity simulations used in this study. I am thankful to Mr Antoine Petiteau for clarifications on the sensitivity response of the European mission of LISA.

\section*{References}
\scriptsize{\bibliography{ref-list}}

\providecommand{\newblock}{}
\begin{thebibliography}{10}
\expandafter\ifx\csname url\endcsname\relax
  \def\url#1{{\tt #1}}\fi
\expandafter\ifx\csname urlprefix\endcsname\relax\def\urlprefix{URL }\fi
\providecommand{\eprint}[2][]{\url{#2}}

\bibitem{Nandra:2009br}
Nandra K, Aird J, Alexander D, Ballantyne D, Barcons X {\em et~al.\/} 2009  *
  Brief entry * (\textit{Preprint} \eprint{0903.0547})

\bibitem{Somerville:2008ch}
Somerville R~S 2008  * Brief entry * (\textit{Preprint} \eprint{0808.1254})

\bibitem{AmaroSeoane:2009ui}
Amaro-Seoane P and Santamaria L 2010 {\em Astrophys.J.\/} {\bf 722} 1197--1206
  (\textit{Preprint} \eprint{0910.0254})

\bibitem{BCW05}
{Berti} E, {Cardoso} V and {Will} C~M 2006 {\em Phys.~Rev.~D\/} {\bf 73}
  064030--+ (\textit{Preprint} \eprint{gr-qc/0512160})

\bibitem{PhysRevD.85.024018}
Kamaretsos I, Hannam M, Husa S and Sathyaprakash B~S 2012 {\em Phys. Rev. D\/}
  {\bf 85}(2) 024018
  \urlprefix\url{http://link.aps.org/doi/10.1103/PhysRevD.85.024018}

\bibitem{Berti:2007a}
Berti E, Cardoso J, Cardoso V and Cavagli\'a M 2007 {\em Phys. Rev.\/} {\bf
  D76} 104044 (\textit{Preprint} \eprint{0707.1202})

\bibitem{Nelemans:2003ha}
Nelemans G, Yungelson L and Portegies~Zwart S 2004 {\em
  Mon.Not.Roy.Astron.Soc.\/} {\bf 349} 181

\bibitem{VanDenBroeck:2010fp}
Van Den~Broeck C, Trias M, Sathyaprakash B and Sintes A 2010 {\em Phys.Rev.\/}
  {\bf D81} 124031 (\textit{Preprint} \eprint{1001.3099})

\bibitem{Berti:2009kk}
Berti E, Cardoso V and Starinets A~O 2009 {\em Class.\ Quant.\ Grav.\/} {\bf
  26} 163001 (\textit{Preprint} \eprint{0905.2975})

\bibitem{Kokkotas:1999bd}
Kokkotas K~D and Schmidt B~G 1999 {\em Living Rev.Rel.\/} {\bf 2} 2
  (\textit{Preprint} \eprint{gr-qc/9909058})

\bibitem{Baker:2008d78}
Baker J~G, Boggs W~D, Centrella J, Kelly B~J, McWilliams S~T and van Meter J~R
  2008 {\em Phys.Rev.\/} {\bf D78} 044046 (\textit{Preprint}
  \eprint{0805.1428})

\bibitem{Brugmann:2008zz}
Br{\"u}gmann B {\em et~al.\/} 2008 {\em Phys. Rev.\/} {\bf D77} 024027
  (\textit{Preprint} \eprint{gr-qc/0610128})

\bibitem{Husa:2007hp}
Husa S, Gonz{\'a}lez J~A, Hannam M, Br{\"u}gmann B and Sperhake U 2008 {\em
  Class. Quant. Grav.\/} {\bf 25} 105006 (\textit{Preprint} \eprint{0706.0740})

\bibitem{Berti:2007b}
Berti E, Cardoso V, Gonzalez J, Sperhake U, Hannam M, Husa S and Brugmann B
  2007 {\em Phys. Rev.\/} {\bf D76} 064034 (\textit{Preprint}
  \eprint{gr-qc/0703053v2})

\bibitem{Hannam:2007ik}
Hannam M, Husa S, Sperhake U, Bruegmann B and Gonzalez J~A 2008 {\em
  Phys.Rev.\/} {\bf D77} 044020 (\textit{Preprint} \eprint{0706.1305})

\bibitem{Hannam:2010ec}
Hannam M, Husa S, Ohme F, Muller D and Bruegmann B 2010 {\em Phys.Rev.\/} {\bf
  D82} 124008 (\textit{Preprint} \eprint{1007.4789})

\bibitem{Valluri:2005up}
Valluri M, Ferrarese L, Merritt D and Joseph C~L 2005 {\em Astrophys. J.\/}
  {\bf 628} 137--152 (\textit{Preprint} \eprint{astro-ph/0502493})

\bibitem{Safonova:2009td}
Safonova M and Shastri P 2010 {\em Astrophys. Space Sci.\/} {\bf 325} 47--58
  (\textit{Preprint} \eprint{0910.2551})

\bibitem{Beifiori:2011be}
Beifiori A, Courteau S, Corsini E~M and Zhu Y 2011  (\textit{Preprint}
  \eprint{1109.6265})

\bibitem{Li:2011ik}
Li Y~R, Ho L~C and Wang J~M 2011 {\em Astrophys. J.\/} {\bf 742} 33
  (\textit{Preprint} \eprint{1109.0089})

\bibitem{Rafiee:2011ry}
Rafiee A and Hall P~B 2011 {\em Astrophys. J. Suppl.\/} {\bf 194} 42
  (\textit{Preprint} \eprint{1104.1828})

\bibitem{Burkert:2010ki}
Burkert A and Tremaine S 2010 {\em Astrophys. J.\/} {\bf 720} 516--521
  (\textit{Preprint} \eprint{1004.0137})

\bibitem{Harris:2010zr}
Harris G~L~H and Harris W~E 2010  (\textit{Preprint} \eprint{1008.4748})

\bibitem{Miller:2008fi}
Miller M~C 2009 {\em Class. Quant. Grav.\/} {\bf 26} 094031 (\textit{Preprint}
  \eprint{0812.3028})

\bibitem{Noyola:2011pf}
Noyola E and Baumgardt H 2011  (\textit{Preprint} \eprint{1108.4425})

\bibitem{FregeauIMBH06}
Fregeau J~M, Larson S~L, Miller M~C, O'Shaughnessy R~W and Rasio F~A 2006 {\em
  Astrophys. J.\/} {\bf 646} L135--L138 (\textit{Preprint}
  \eprint{astro-ph/0605732})

\bibitem{1538-3881-135-1-182}
Bash F~N, Gebhardt K, Goss W~M and Bout P~A~V 2008 {\em The Astronomical
  Journal\/} {\bf 135} 182
  \urlprefix\url{http://stacks.iop.org/1538-3881/135/i=1/a=182}

\bibitem{Volonteri:2004cf}
Volonteri M, Madau P, Quataert E and Rees M~J 2005 {\em Astrophys. J.\/} {\bf
  620} 69--77 (\textit{Preprint} \eprint{astro-ph/0410342})

\bibitem{Wainstein}
Wainstein L~A and Zubakov V~D 1962 {\em Extraction of Signals from Noise\/}
  (Englewood Cliffs: Prentice-Hall)

\bibitem{Finn92}
Finn L~S 1992 {\em Phys. Rev. D\/} {\bf 46} 5236

\bibitem{BalSatDhu96}
Balasubramanian R, Sathyaprakash B~S and Dhurandhar S~V 1996 {\em
  Phys.~Rev.~D\/} {\bf 53} 3033 erratum-ibid.~D {\bf 54}, 1860 (1996)
  (\textit{Preprint} \eprint{gr-qc/9508011})

\bibitem{SathyaSchutzLivRev09}
Sathyaprakash B~S and Schutz B~F 2009 {\em Living Rev. Rel.\/} {\bf 12} 2
  (\textit{Preprint} \eprint{arXiv:0903.0338})

\bibitem{Hild:2009ns}
Hild S, Chelkowski S, Freise A, Franc J, Morgado N {\em et~al.\/} 2010 {\em
  Class.Quant.Grav.\/} {\bf 27} 015003 (\textit{Preprint} \eprint{0906.2655})

\bibitem{Filloux:2011eq}
Filloux C, Pacheco J~A~d~F, Durier F and de~Araujo J~C~N 2011
  (\textit{Preprint} \eprint{1108.2638})

\bibitem{Berti06}
Berti E 2006 {\em Classical Quantum Gravity\/} {\bf 23} S785 (\textit{Preprint}
  \eprint{astro-ph/0602470})

\bibitem{0264-9381-20-10-304}
Haehnelt M~G 2003 {\em Classical and Quantum Gravity\/} {\bf 20} S31
  \urlprefix\url{http://stacks.iop.org/0264-9381/20/i=10/a=304}

\bibitem{Rhook:2005pt}
Rhook K~J and Wyithe J~S~B 2005 {\em Mon. Not. Roy. Astron. Soc.\/} {\bf 361}
  1145--1152 (\textit{Preprint} \eprint{astro-ph/0503210})

\bibitem{Erickcek:2006xc}
Erickcek A~L, Kamionkowski M and Benson A~J 2006 {\em Mon. Not. Roy. Astron.
  Soc.\/} {\bf 371} 1992--2000

\bibitem{Micic:2007vd}
Micic M, Holley-Bockelmann K, Sigurdsson S and Abel T 2007 {\em Mon. Not. Roy.
  Astron. Soc.\/} {\bf 380} 1533

\bibitem{Wen:2011xc}
Wen Z~L, Jenet F~A, Yardley D, Hobbs G~B and Manchester R~N 2011 {\em
  Astrophys. J.\/} {\bf 730} 29 (\textit{Preprint} \eprint{1103.2808})

\bibitem{deFreitasPacheco:2006bh}
de~Freitas~Pacheco J~A, Filloux C and Regimbau T 2006 {\em Phys. Rev.\/} {\bf
  D74} 023001

\bibitem{Menou:2001hb}
Menou K, Haiman Z and Narayanan V~K 2001 {\em Astrophys.J.\/} {\bf 558}
  535--542 (\textit{Preprint} \eprint{astro-ph/0101196})

\bibitem{AmaroSeoane:2006py}
Amaro-Seoane P and Freitag M 2006 {\em Astrophys. J.\/} {\bf 653} L53--L56
  (\textit{Preprint} \eprint{astro-ph/0610478})

\bibitem{rates:2010cf}
Abadie J {\em et~al.\/} (LIGO Scientific Collaboration) 2010
  (\textit{Preprint} \eprint{arXiv:1003.2480})

\bibitem{Mandel:2009dn}
Mandel I, Gair J~R and Miller M~C 2009  (\textit{Preprint} \eprint{0912.4925})

\bibitem{White:2011qf}
White D~J, Daw E and Dhillon V 2011 {\em Class.Quant.Grav.\/} {\bf 28} 085016
  (\textit{Preprint} \eprint{1103.0695})

\end{thebibliography}
\end{document}